\documentclass[12pt,utf8,a4paper,english,oneside]{article}

\usepackage[utf8]{inputenc}
\usepackage[OT1]{fontenc} 

\usepackage{graphicx}

\usepackage{amsmath}
\usepackage{amsfonts}
\usepackage{amssymb}
\usepackage{amsbsy}
\usepackage{dsfont}

\usepackage{latexsym}
\usepackage{array}

\usepackage{hyperref}
\hypersetup{
    colorlinks=true,
    linkcolor=blue,
    citecolor=black,
    urlcolor=blue,
}

\usepackage{float}
\usepackage{textcomp}
\usepackage{xcolor}
\usepackage{times}
\usepackage{authblk}

\newcommand{\mb}{\mathbf}
\newcommand{\mbb}{\mathbb}
\newcommand{\mc}{\mathcal}
\newcommand{\bs}{\boldsymbol}
\newcommand{\bc}{\begin{center}}
\newcommand{\ec}{\end{center}}
\newcommand{\nit}{\noindent}

\newcommand{\beq}{\begin{equation}}
\newcommand{\eeq}{\end{equation}}
\newcommand{\beqa}{\begin{eqnarray}}
\newcommand{\eeqa}{\end{eqnarray}}
\newcommand{\beqan}{\begin{eqnarray*}}
\newcommand{\eeqan}{\end{eqnarray*}}
\newcommand{\bit}{\begin{itemize}}
\newcommand{\eit}{\end{itemize}}

\DeclareMathOperator{\diag}{diag}
\begin{document}

%\title{Effective Connectivity deduced with Graph Convolutional Networks from rs-fMRI Data}
%\title{A Multi-modal Graph Neural Network Approach for Effective Brain Connectivity}
%\title{A Graph Neural Network Framework for Effective Brain Connectivity}
\title{A Graph Neural Network Framework for Causal Inference in Brain Networks}

\author[1,2]{S. Wein \thanks{Simon.Wein$@$ur.de}}
\author[2]{W. Malloni}
\author[3]{A.~M.~Tom\'e}
\author[4]{S. M. Frank}
\author[2]{G.-I. Henze}
\author[2]{S. Wüst}
\author[2]{M. W. Greenlee}
\author[1]{E. W. Lang}

\affil[1]{\normalsize CIML, Biophysics, University of Regensburg, Regensburg, Germany}
\affil[2]{Experimental Psychology, University of Regensburg, Regensburg, Germany}
\affil[3]{IEETA/DETI, Universidade de Aveiro, Aveiro, Portugal}
\affil[4]{Department of Cognitive, Linguistic,and Psychological Sciences, Brown University, Providence, RI 02912, USA}

\maketitle

%%%%%%%%%%%%%%%%%%%%%%%%%%%%%%%%
\begin{abstract}
A central question in neuroscience is how self-organizing dynamic interactions in the brain emerge on their relatively static structural backbone. Due to the complexity of spatial and temporal dependencies between different brain areas, fully comprehending the interplay between structure and function is still challenging and an area of intense research. In this paper we present a graph neural network (GNN) framework, to describe functional interactions based on the structural anatomical layout. A GNN allows us to process graph-structured spatio-temporal signals, providing a possibility to combine structural information derived from diffusion tensor imaging (DTI) with temporal neural activity profiles, like observed in functional magnetic resonance imaging (fMRI). Moreover, dynamic interactions between different brain regions learned by this data-driven approach can provide a multi-modal measure of causal connectivity strength. We assess the proposed model's accuracy by evaluating its capabilities to replicate empirically observed neural activation profiles, and compare the performance to those of a vector auto regression (VAR), like typically used in Granger causality. We show that GNNs are able to capture long-term dependencies in data and also computationally scale up to the analysis of large-scale networks. Finally we confirm that features learned by a GNN can generalize across MRI scanner types and acquisition protocols, by demonstrating that the performance on small datasets can be improved by pre-training the GNN on data from an earlier and different study. We conclude that the proposed multi-modal GNN framework can provide a novel perspective on the structure-function relationship in the brain. Therewith this approach can be promising for the characterization of the information flow in brain networks.    
\end{abstract}

%%%%%%%%%%%%%%%%%%%%%%%%%%%%%%%%%


%%%%%%%%%%%%%%%%%%%%%%%%%%%%%%%%%%
%%%%%%%%%%%%%%%%%%%%%%%%%%%%%%%%%%
%%%%%%%%%%%%%%%%%%%%%%%%%%%%%%%%%%
\section{Introduction}

Brain connectivity comes in different flavors, either resting on the structural anatomical layout, as derived from diffusion tensor imaging (DTI) or based on temporally resolved activity patterns, like observed in functional MRI (fMRI) \cite{Lang2012}. White matter tracks reconstructed from DTI provide a foundation for structural connectivity (SC) and can be used to quantify the (static) anatomical connection strength between brain regions. On the other hand fMRI enables us to map out dynamic neural activity distributions across the brain, whereas the coherence of fluctuations is usually referred to as functional connectivity (FC). Intuitively one might follow the paradigm \textit{"structure determines function"}, but it has been shown that the relationship between brain structure and function is quite complex and still a focus of intense research \cite{Deco2012, Hermundstad2013, Messe2014, Abdelnour2014, Bettinardi2018}. For instance, brain regions with robust SC usually show also high FC, but the inverse is not necessarily true \cite{Honey2009}. While FC is a statistical measure with no information concerning the directionality of the relation, effective connectivity and directed functional connectivity measures try to infer causal dependencies in functional imaging data \cite{Friston2013}. Thus connectivity measures derived from different modalities can provide distinct, but complementary aspects of brain connectivity \cite{Amico2018, Xue2015, Chu2018}. Still, studying their relations is challenging mainly due to the complex spatio-temporal dependencies and inherent difficulty in long term forecasting.

In this paper we propose a data driven model, which combines information from fMRI and DTI to infer causal dependencies between brain regions. Temporal activity patterns of neuron pools, interconnected by the spatial anatomical layout, can be interpreted as time-varying graph structured signals. For such applications, graph neural networks (GNN) have shown to be useful, providing a possibility to process data with graph-like properties in the framework of artificial neural networks (ANN) \cite{Wu2019}. Motivated by their success in computer vision \cite{Fukushima1987, Lecun2015}, convolution operations were recently extended to the graph domain \cite{Bruna2014, Defferrard2016}. Learning such convolution filters in ANN enables us to capture inherent spatial dependencies in the non-Euclidean geometry of graphs, which are used in our context to integrate spatial relations of brain networks, based on their structural anatomical connections. Further, temporal dependencies in a dynamic system can be acquired by recurrent neural networks (RNNs) that have proven to be well suited for processing data with sequential structure. In our study, RNNs learn temporal characteristics of brain dynamics, like those observed in resting-state fMRI. A certain type of GNN architecture denoted as \textit{diffusion convolution recurrent neural network} (DCRNN), \cite{Li2018} provides the possibility to integrate spatio-temporal information of graph-structured signals. By combining fMRI with DTI data, the idea is to replicate brain dynamics more accurately, to get an improved understanding of functional interactions between brain regions, which are physically constrained by their structural backbone \cite{Deco2017}.

Causal relationships between brain regions can be revealed by directed functional connectivity and effective connectivity. Two prominent and distinct approaches have been established in recent years therefore \cite{Friston2013}. The first one is based on a simple idea taken up by the British econometrician Clive Granger \cite{Granger1969}. If one event $A$ causes another event $B$, then $A$ would precede $B$, and information on the occurrence of $A$ should contribute to the prediction of the occurrence of event $B$. Such temporal dependencies between multivariate processes are typically described in the framework of a multivariate vector auto regressive (VAR) model, building a foundation for Granger causality (GC). By trying to make accurate predictions of temporal neural profiles, GC tests if adding information about neural activity in brain region $B$ helps to improve the prediction of the activity in region $A$ (and vice versa). This provides an exploratory measure for directed causal dependencies between segregated brain areas. 

The second popular approach is methodologically different: Dynamic causal modeling  (DCM) relies on a mechanistic input-state-output model of neuron pools, describing the effective connectivity strength between brain areas \cite{Friston2003}. Experimental conditions and stimuli are encoded in input functions, and the model output can be related to empirically observed electromagnetic or hemodynamic responses. In a Bayesian framework, effective couplings of neural populations are estimated, providing a neurophysiological perspective on causal relationships between different regions in the brain. However due to its relatively high computational complexity, the analysis with DCM is usually limited to a few pre-defined regions in the brain only, what could neglect relevant components for the analysis \cite{Daunizeau2009}. 

Here we present a data-driven machine learning approach that combines structural and functional information of neuron pools in a predictive framework for brain dynamics. By studying spatio-temporal dependencies between brain areas which were learned by the DCRNN model from DTI and fMRI data, we deduce the information flow between segregated areas in the brain. This provides us with a multi-modal data-driven perspective on causal relationships within brain networks. We compare the predictive capabilities of a typical VAR model to those of a DCRNN model, and show that this machine learning approach can successfully account for long-term and non-linear relationships in data. We conclude that such data-driven methods inspired by graph signal processing can be promising candidates for modeling brain dynamics, foremost due to their improved accuracy in replicating empirically observed data. Moreover, a greater neurophysiological plausibility results, because neural interactions are constrained by their anatomical substrates in this model. In contrast to classical DCM, the DCRNN also naturally scales to large networks by learning localized filters on the graph structure \cite{Defferrard2016}, what opens the possibility for an exploratory analysis of whole brain networks.  

Usually for a good performance more complex machine learning models require a larger amount of data, but it is not economical in MRI to perform studies with very large sample sizes. To account for these issues we demonstrate that also in our context transfer learning \cite{Pan2010} can enhance the model accuracy of small datasets. We pre-train the DCRNN on a large-scale dataset of $100$ resting-state fMRI (rs-fMRI) sessions provided by the Human Connectome Project \cite{VanEssen2013} (HCP). We then show that the pre-trained model considerably improves the predictive performance on a smaller independent dataset of $10$ sessions compared to standard training. This points to the ability of the DCRNN to generalize across scanner types and acquisition protocols to a certain extent, enabling the possibility for transfer learning.

Finally the integrative framework of anatomical and functional neuroimaging data can help to better understand the general relation between brain structure and function. By modeling localized interactions on the anatomical substrate, this approach allows us to reconstruct the amount of information on activity distributions that occurs in structurally connected brain regions. While many current approaches focus on predicting only the coherence patterns of the activity in brain regions (FC) from their SC \cite{Becker2018, Surampudi2018, Saggio2016, Liang2017, Deligianni2016}, we present a framework that directly replicates observed neural activity profiles, thereby relying on information from SC.   

%%%%%%%%%%%%%%%%%%%%%%%%%%%%%%%%%%
%%%%%%%%%%%%%%%%%%%%%%%%%%%%%%%%%%
%%%%%%%%%%%%%%%%%%%%%%%%%%%%%%%%%%

\section{Results} \label{sec:results}

\subsection{Model description.}

\begin{figure}[!htb]
\bc
\makebox[\textwidth][c]{\includegraphics[width=1.25\textwidth]{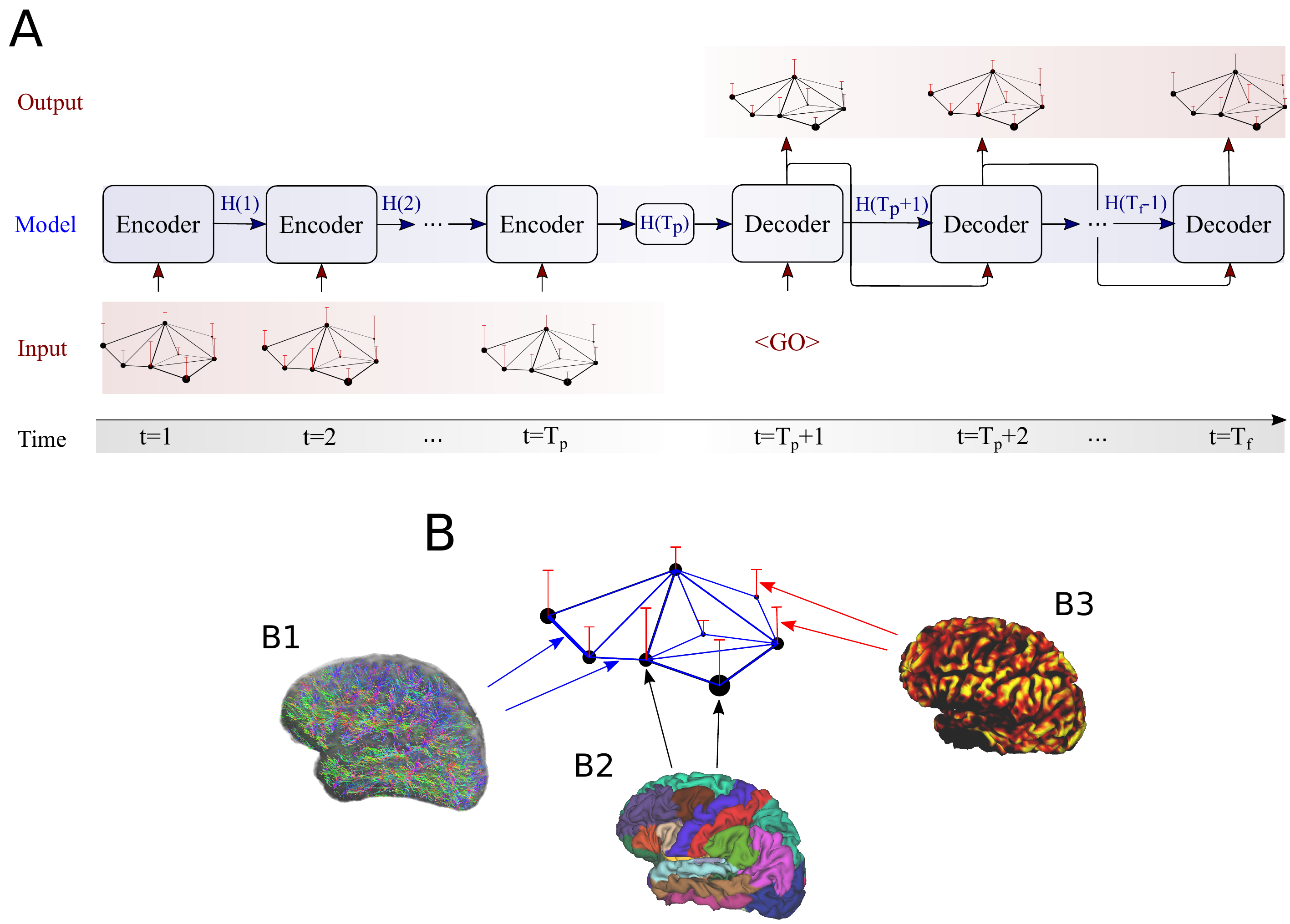}}
\ec 
\caption{An overview of the DCRNN model. The model consists of an encoder and decoder (A), modified to process graph structured signals (B). In our context, vertices (nodes) $\mc{V}, |\mc{V}| = N$ of the graph $\mc{G}$ are defined as $N$ brain regions, derived from an atlas (B2). Structural connections between brain regions are derived from DTI, quantifying the strength of edge connections in the graph (B1). The signal on the graph $\mb{x}(t)$ at a certain time point $t$ is the average BOLD signal in brain regions/nodes, obtained by the fMRI measurement at time $t$ (B3). The encoder (A) receives an input sequence $[\mb{x}(1),\ldots,\mb{x}(T_p)]$, and iteratively updates its hidden state $H(t)$. The final encoder state $H(T_p)$ is passed to the decoder part, which learns to recursively predict the output sequence of graph signals $[\mb{x}(T_p+1),\ldots,\mb{x}(T_f)]$ in the future. During testing and validation, the decoder uses its own outputs as inputs, to generate the subsequent output. The first input of the decoder ($<GO>$ symbol) is simply a vector of zeros.}
\label{fig:DCRNN}
\end{figure}

In this study we use the DCRNN model \cite{Li2018} architecture to explore the spatio-temporal relationships of brain dynamics in resting state fMRI. An overview of the model structure is provided in figure \ref{fig:DCRNN}. To learn the temporal dependencies of the BOLD signal, recurrent neural networks (RNNs) with sequence-to-sequence learning are employed \cite{Sutskever2014}. In such an architecture the encoder network maps information from an input sequence into a hidden representation, which is used by the decoding part to sequentially generate outputs, based on this encoded information. In the context of brain dynamics, the input sequence corresponds to measurements of the BOLD signal $\mb{x}(t)  \in \mbb{R}^{N}$ in $N$ brain regions at $T_p$ time points, while the objective is to predict the signal at $T_f$ subsequent time points. 

In addition to temporal, also spatial dependencies between brain regions are incorporated via diffusion convolution operations \cite{Li2018}. Consider the network of regions of interest (ROIs) as a {\em graph} $\mc{G} = (\mc{V}, \mc{E},\mb{A}_w)$, where $\mc{V}, |\mc{V}| = N$ denotes a set of vertices (nodes), $\mc{E}$ represents a set of edges and $\mb{A}_w \in \mbb{R}^{N \times N}$ is a {\em weighted adjacency matrix}. The latter represents the spatial connectivity of the nodes, i.e. the ROIs on the neuronal network, which are adjacent to each other, i.e. connected by an edge. Also the weights result from DTI, reflecting the axonal connection strength between the connected regions. Goal of the DCRNN model is to learn a function $h(...)$ which maps $T_p$ past activity states $\mb{x}(t)$, to $T_f$ future states:   

\beq \label{eq:state_mapping}
[\mb{x}(t-T_p+1), \ldots, \mb{x}(t); \mc{G}] \xrightarrow{h(...)} [\mb{x}(t+1), \ldots, \mb{x}(t+T_f)]
\eeq 
The encoder, as well as the decoder of the DCRNN consist of gated recurrent units \cite{Chung2014}, modified with graph convolutions \cite{Defferrard2016}, and for training the model scheduled sampling was applied \cite{Bengio2015}. A detailed description of the model architecture is provided in section \ref{sec:methods}.

%%%%%%%%%%%%%%%%%%%%%%%%%%%%%%%%%%
%%%%%%%%%%%%%%%%%%%%%%%%%%%%%%%%%%

\subsection{Data description.} \label{sec:data_descr}
For the first part of our evaluation, resting-state fMRI data from the \textit{S1200 release} provided by the \textit{Human Connectome Project} \cite{VanEssen2013} (HCP) was employed \cite{Glasser2013}. Further the multi-model parcellation proposed by Glasser et. al \cite{Glasser2016} was applied to divide each hemisphere into 180 segregated regions. The BOLD signal in each region was averaged, so for each resting state session, $N=360$ time courses were obtained. During each session $T=1200$ images were acquired, so the data can be arranged in a matrix $\mb{X} \in \mbb{R}^{N \times T}$. For the following analysis, we filter the data with a $0.04-0.07 Hz$ narrow band bandpass filter, as it has shown to be reliable and functionally relevant for gray matter activity \cite{Glearean2012, Bruckner2009, Deco2017, Biswal1995, Achard2006}. We additionally present results in supplement \hyperref[sec:supplement_1]{I}, when employing a more liberal bandpass filter with cutoff frequencies between $0.02-0.09 Hz$.

The input and output (label) samples for the DCRNN model were generated from the data in $\mb{X}$, by defining windows of length $T_p$ to obtain input sequences of neural activity states $[\mb{x}(t-T_p+1), \ldots, \mb{x}(t)]$, and respective target sequences $[\mb{x}(t+1), \ldots, \mb{x}(t+T_f)]$ of length $T_f$. The index $t$ was propagated through each resting-state fMRI session, so in total $T - T_p - T_f + 1$ input-output pairs were generated per session. The first $80\%$ were used for training the DCRNN model, $10\%$ for validation, and the last $10\%$ for testing. In total $4$ resting-state fMRI sessions from $25$ different subjects were employed for the evaluations. The input and output length was chosen to be $T_p = T_f = 30$, what would correspond to a time span of roughly $22\ s$, based on the sampling with a repetition time $TR = 0.72\ s$ \cite{Ugurbil2013}. But note that in general the sequence-to-sequence architecture employed would be able to deal with arbitrary input and output signal lengths \cite{Sutskever2014}.

In addition to temporal brain dynamics, also structural information was incorporated into the model, described by the anatomical connection strength between brain regions deduced from DTI. Therefore the DTI dataset provided in the \textit{S1200 release} \cite{Glasser2013} was further processed by employing multi-shell, multi-tissue constrained spherical deconvolution \cite{Jeurissen2014}, implemented in the \textit{MRtrix3} software package \cite{Tournier2019}. White matter tracktography was performed to estimate whole brain structural connectivity between the $N=360$ regions of the multi-modal parcellation atlas \cite{Glasser2016}. The reconstructed anatomical connections define the edge strength in the graph adjacency matrix $\mb{A}_w \in \mbb{R}^{N \times N}$. A more detailed description of the datasets and preprocessing involved can be found in section \ref{sec:HCP_data}.

%%%%%%%%%%%%%%%%%%%%%%%%%%%%%%%%%%
%%%%%%%%%%%%%%%%%%%%%%%%%%%%%%%%%%

\subsection{Model performance.} \label{sec:performance}

In a first step we assess the capabilities of the DCRNN model to learn temporal activity patterns in neuron pools, and their relationships across the spatial layout. As a first baseline we compare the DCRNN to the performance of a linear vector autoregressive (VAR) model \cite{Granger1969}, further described in section \ref{sec:VAR}. A common way to estimate causal relations between different regions of interest (ROIs) in a brain network, is to fit a multivariate VAR model to neural temporal activity patterns, like those observed in different neuroimaging modalities \cite{Friston2013, Barnett2013, Seth2015}. Evaluating the fitted VAR allows us to infer, if one spatial brain region, contains additional information about future activity profiles of other regions, indicating a causal dependency between them. The accuracy in replicating empirically observed neural activity profiles can indicate how well a model has learned the underlying process of neural dynamics, including the interactions and dependencies among brain regions. In this comparison we incorporated two different optimization methods for the estimation of the VAR coefficients. The first one employs an ordinary least squares (OLS) fit on the neural activity timecourses $\mb{x}(t)$ from individual rs-fMRI sessions \cite{Hamilton1994}. The second approach, in analogy to the DCRNN, follows a gradient-descent based optimization \cite{Kiefer1952} on the windowed neural activity samples as described in section \ref{sec:data_descr}. For this evaluation we rely on the latter one, as it could improve the performance of the VAR, as outlined in section \ref{sec:VAR}.

\begin{figure}[!htb]
\bc
\makebox[\textwidth][c]
{\includegraphics[width=1.2\textwidth]{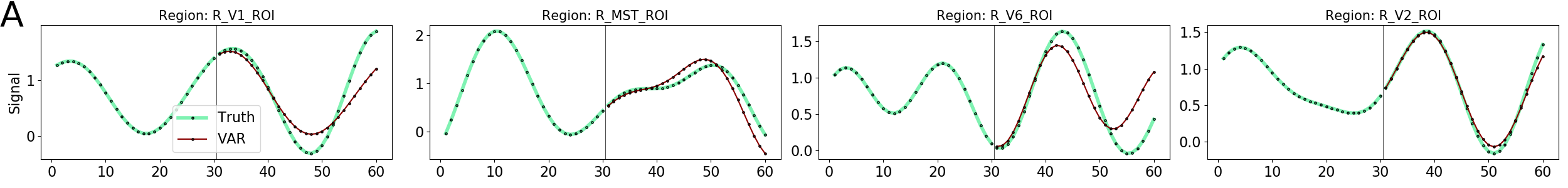}
}

\vspace{4mm}

\makebox[\textwidth][c]
{\includegraphics[width=1.2\textwidth]{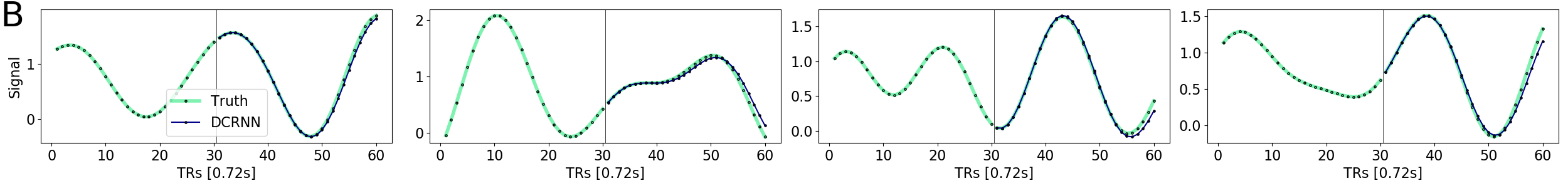}
}

\vspace{1mm}

\makebox[\textwidth][c]
{\includegraphics[width=1\textwidth]{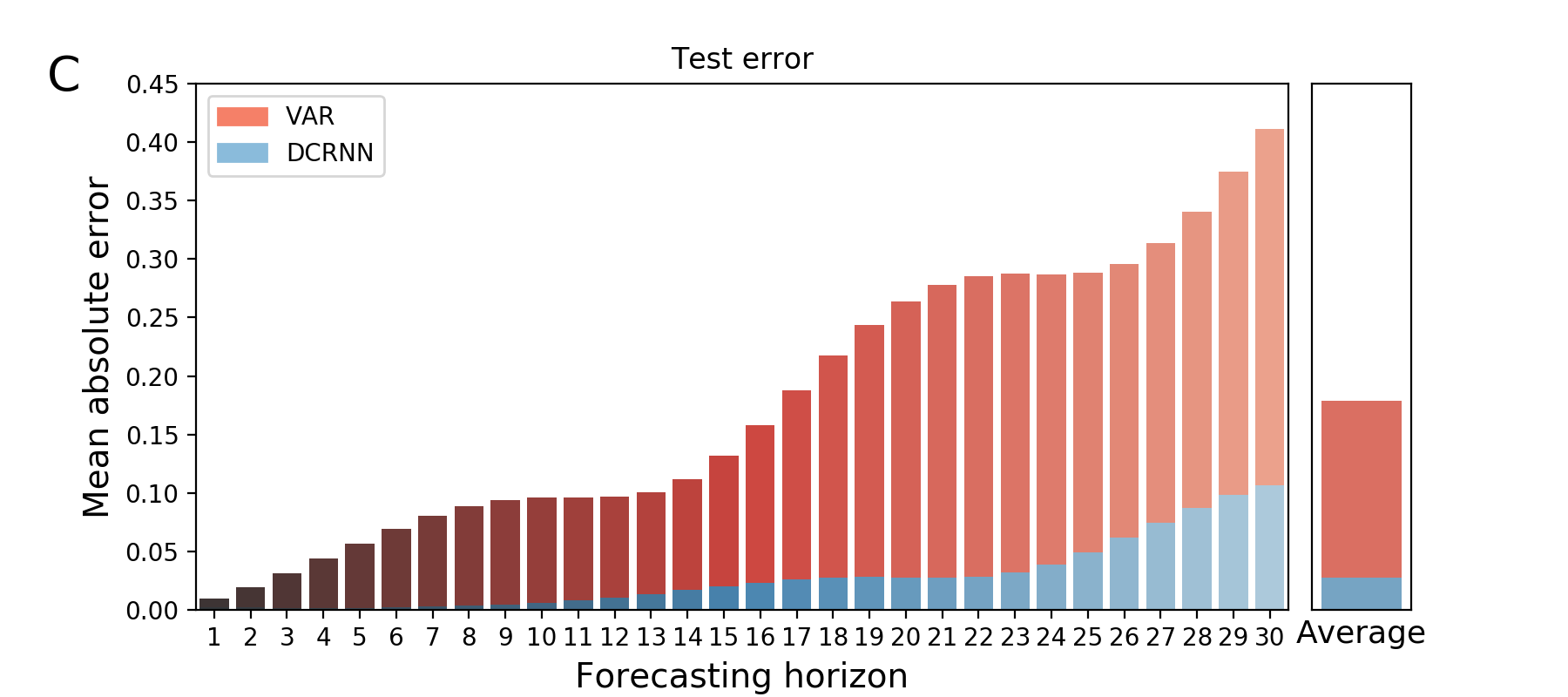}
}
\ec 
\caption{The figure illustrates the prediction accuracy of a VAR model (A) in comparison to the DCRNN (B). The true BOLD signal in these 4 ROIs is marked green, while predictions of the VAR are highlighted in red, and for the DCRNN in blue. The first $30 \,\,TRs$ of BOLD signal were used as the model inputs, and the goal was to predict the subsequent $30 \,\,TRs$. This illustrative example was chosen to represent the whole test set, the prediction error of the VAR model on this sample is $0.169$, and as such slightly below average, while the error of the DCRNN is with $0.037$ higher than its average. Below the average MAE over all samples in the test set is illustrated, in dependence of the forecasting horizon (C). On the right side in (C) the average of all horizons is shown.}
\label{fig:comparison}
\end{figure}

A representative example of the predictive accuracy of both approaches is shown in figure \ref{fig:comparison}, as well as their average performance on the complete testing data set. Figure \ref{fig:comparison} (A) illustrates that a linear VAR model can generate in a few cases also correct long term predictions, but most often after $10 \,\, TRs$ ($\approx 7s$) the error starts to accumulate and the predictions become less accurate. The predictions of the DCRNN (figure \ref{fig:comparison} (B)) remain stable over much longer forecasting horizons, and the average mean absolute error $MAE=0.0279$ is considerably lower than the $MAE=0.1786$ of the VAR.

To further verify the difference in the prediction accuracy, the equivalent evaluation, using a more liberal frequency filtering within the $0.02-0.09 Hz$ range, is provided in supplement \hyperref[sec:supplement_1]{I}. Furthermore in supplement \hyperref[sec:supplement_2]{II} we evaluate the different approaches employing the volumetric AAL parcellation \cite{Tzourio2002} and performing an alternative method for reconstructing the anatomical connectivity \cite{Behrens2007}. 

%%%%%%%%%%%%%%%%%%%%%%%%%%%%%%%%%%
%%%%%%%%%%%%%%%%%%%%%%%%%%%%%%%%%%

\subsection{Impact of spatial modeling.}
\label{sec:impact_structure}

\begin{figure}[!htb]
\centering\makebox[\textwidth][c]
{
  \begin{minipage}[b]{0.52\textwidth}
  \includegraphics[width=1\textwidth]{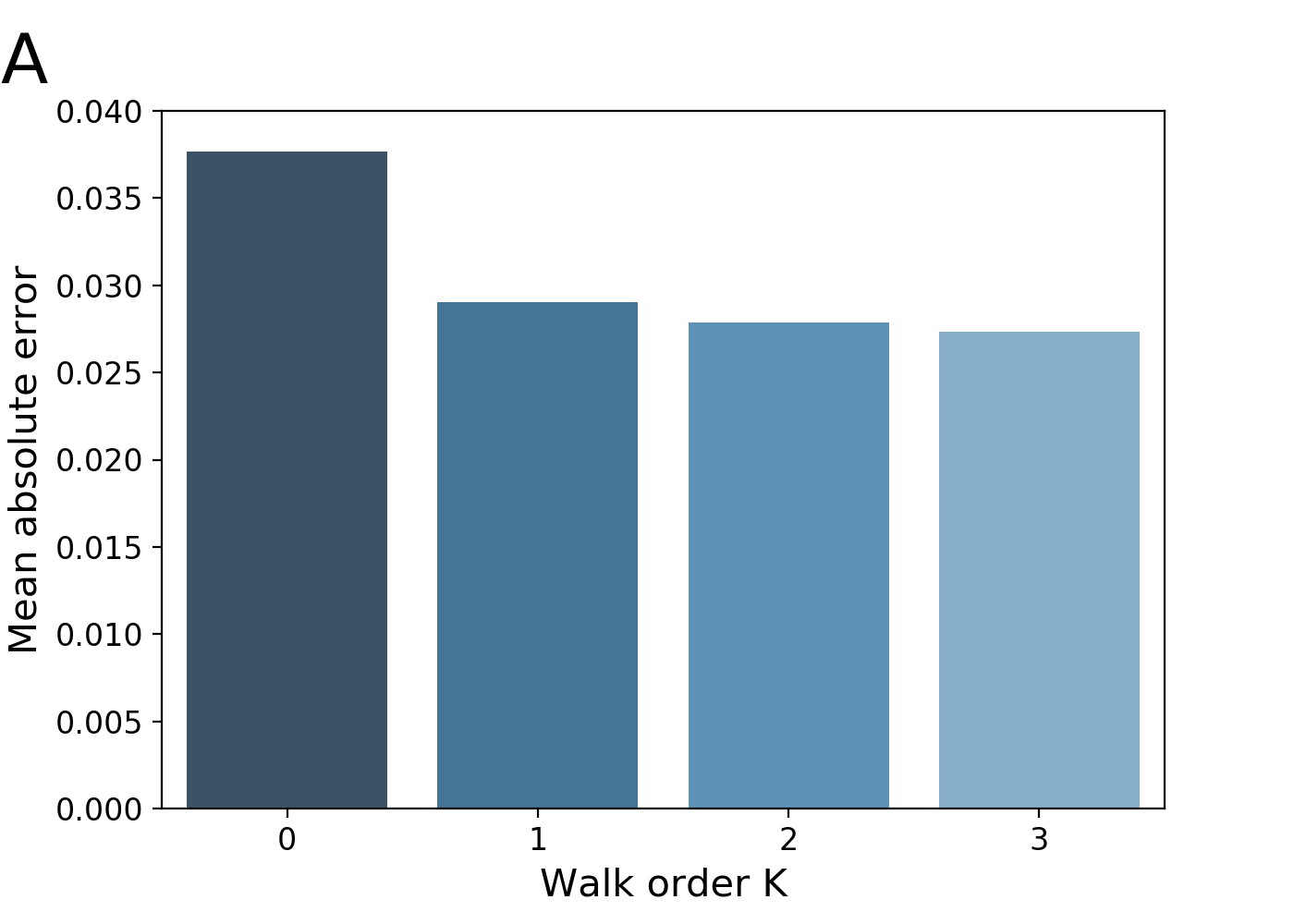}  
  \end{minipage}
  
  \hspace{-2mm}
  
  \begin{minipage}[b]{0.52\textwidth}
  \includegraphics[width=1\textwidth]{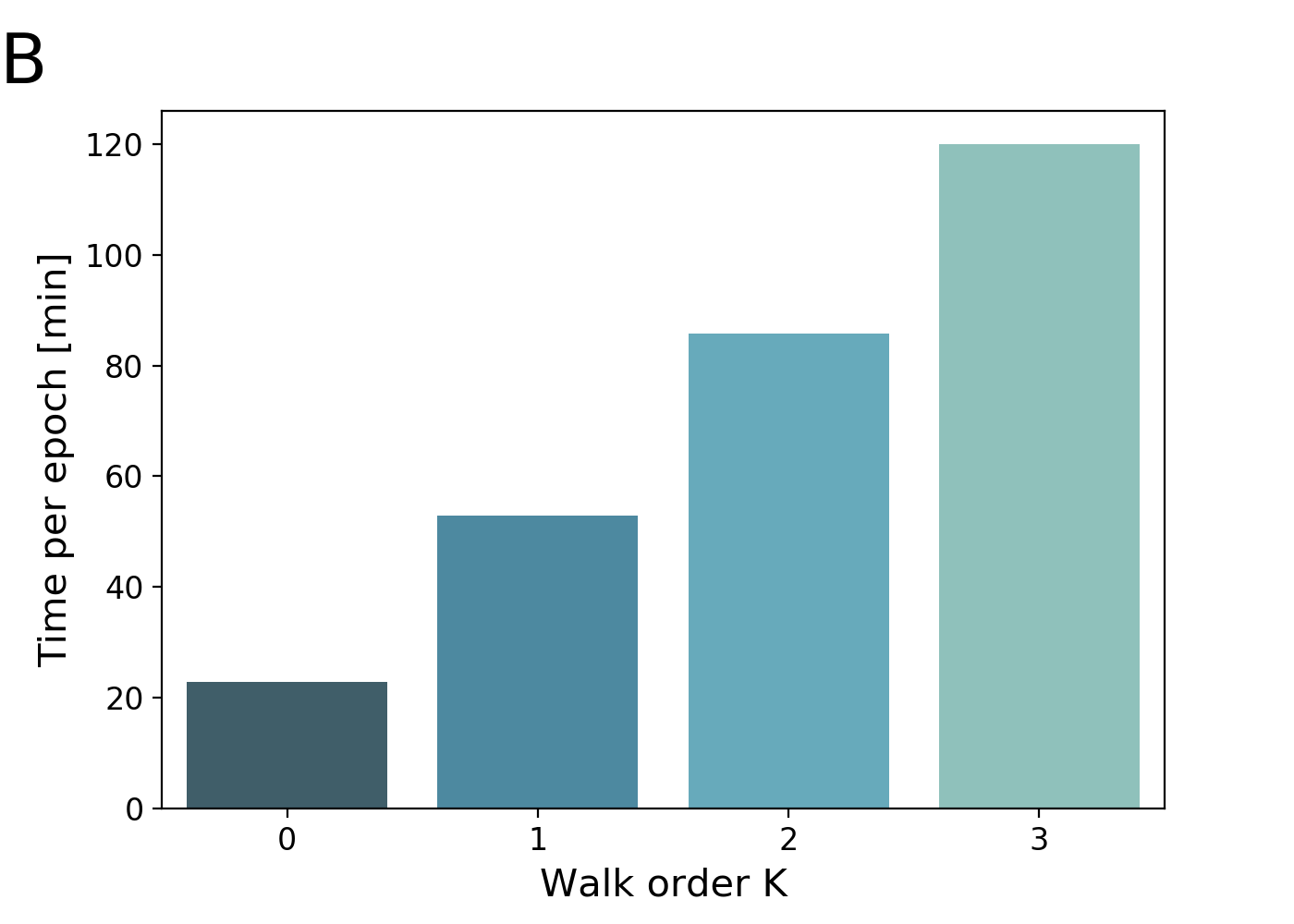}
  \end{minipage}  
}

\includegraphics[width=1\textwidth]{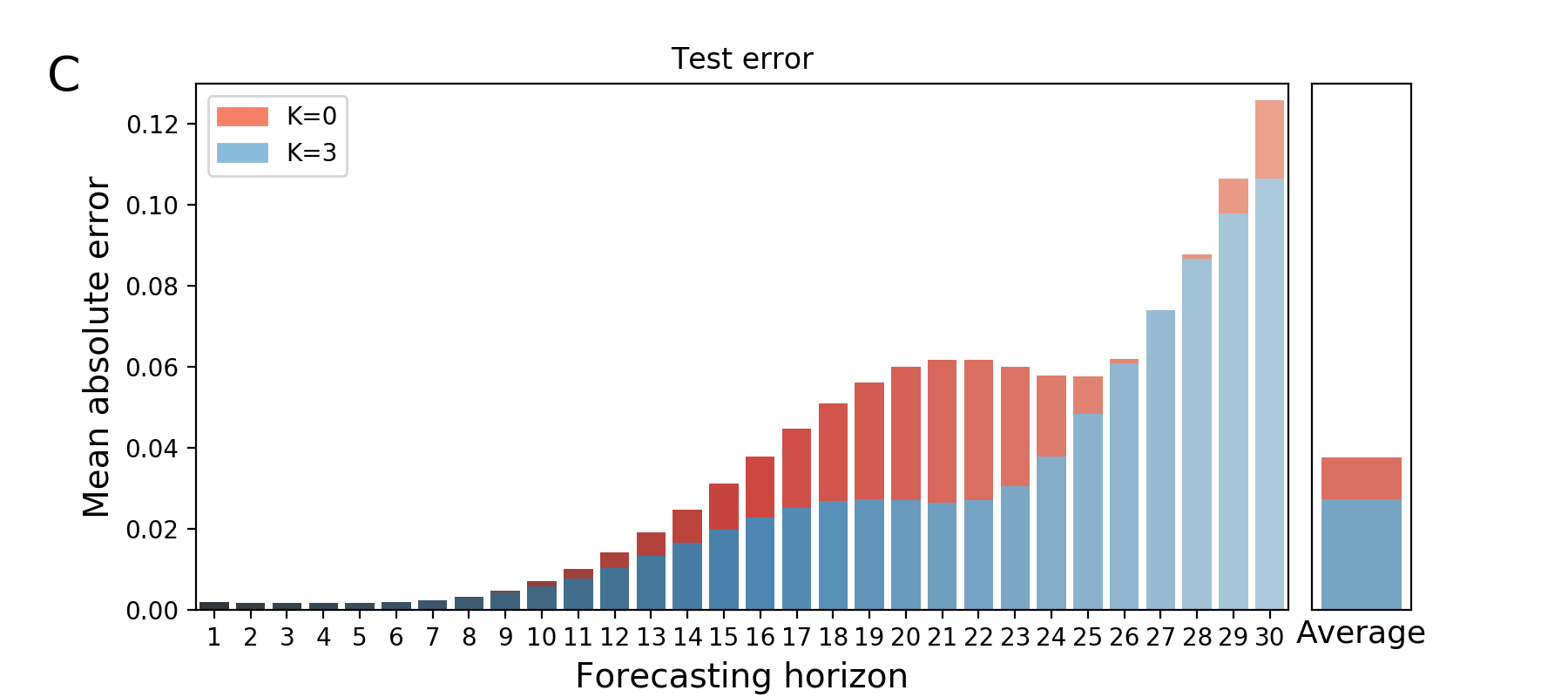}
  
\caption{This figure depicts the effect of structural modeling on the prediction accuracy. In (A) the test MAE in dependence on included walk order $K$ is shown, while (B) demonstrates the impact of $K$ on the computational load per epoch. A more detailed comparison of the MAE on the forecasting horizon when employing filters with order $K=0$ and $K=3$ is illustrated in (C).}
\label{fig:walkorder}
\end{figure}

For this application of the DCRNN model, the anatomical connectivity was used to characterize spatial relations between nodes in the brain network, shaping the transition of activity between brain regions. To illustrate that the DCRNN indeed has learned relevant spatial interactions between different ROIs, we evaluate this recurrent neural network model, without employing graph (diffusion) convolution layers. This restriction considers only self-couplings (filters of order $K=0$) of nodes on the structural graph. Figure \ref{fig:walkorder} (A) shows the test MAE in dependence of the incorporated walk order $K$. The increase in computational time per epoch in dependence of included transition orders $K$ is depicted in figure (B). A more detailed comparison of the prediction MAE between the sequence-to-sequence model without graph convolutions ($K=0$), and including spatial transitions up to order $K=3$ is illustrated in figure \ref{fig:walkorder} (C).

These results show, that the vast amount of the information about future activity in one region comes from the region itself. But by including first order transitions on the structural network ($K=1$) the error can already be decreased by $25 \%$. Filters of higher orders $K=2,3$ only slightly improve the predictions further, but the computational load increases linearly with order $K$, as shown in subfigures (A) and (B). The role of such transitions within the anatomical network can tell us something about the general structure-function relationship in the human brain, by pointing out how much information about functional dynamics comes from structurally connected regions. The comparison between $K=0$ and $K=3$ shows, that roughly up to $27 \%$ of the predictive performance can be attributed to information from regions that are anatomically connected with each other. 

\clearpage

%%%%%%%%%%%%%%%%%%%%%%%%%%%%%%%%%%
%%%%%%%%%%%%%%%%%%%%%%%%%%%%%%%%%%
\subsection{Causal connectivity.}  \label{sec:EC}

In this section the objective is to study the principle of information passing between different ROIs the DCRNN has learned from the neuroimaging data. As shown in the previous section \ref{sec:impact_structure}, propagating information on the anatomical network can improve the predictions of the temporal evolution of the BOLD signal, displaying a dependence among structurally connected brain regions. Now to derive a measure of causal connectivity strength, by following the idea of Granger \cite{Granger1969}, the goal is to reconstruct how information about the activity in ROI \textit{A} contributes to the prediction of the activity in ROI \textit{B}. To reveal relationships inside the data by directly looking at the learned parameters is often challenging when ANN models become more complex. One simple strategy used to account for this problem is to induce perturbations in the models input space and then observe how these perturbations are propagated to the models outputs \cite{Zeiler2013, Samek2017}.

In our context, the DCRNN first learns a function $h(...)$, mapping the original input sequences of neural activity states $ [\mb{x}(t-T_p+1), \ldots, \mb{x}(t)] $ to a predicted output sequences of future states $[\mb{\hat{x}}(t+1), \ldots, \mb{\hat{x}}(t+T_f)]$. Then the information about the activity in a ROI $n'$ is removed, by simply replacing the values $x_{n'}(t)$ in the input sequence with the mean value of the data distribution $\bar{x}_{n'}(t) = 0$. Next the input sequence with the artificial perturbation in $n'$ is projected by the model $h(...)$ to an output sequence $[\mb{\hat{x}'}(t+1), \ldots, \mb{\hat{x}'}(t+T_f)]$. Finally the differences of the models predictions $\mb{\hat{x}}'(t)$ with the perturbation in the input space in ROI $n'$, and the predictions $\mb{\hat{x}}(t)$ with the original input can be compared. A measure of influence $\mb{I}(n') \in \mbb{R}^N $ of the information in ROI $n'$ on the predictions in other ROIs can then be defined as:

\beq
	I_n(n') = \frac{1}{S} \sum_{s=0}^{S} \frac{1}{T_f} \sum_{t=0}^{T_f} | \mb{\hat{x}}_n^{(s)}(t) - {\mb{\hat{x}}_n}'^{(s)}(t) |
\eeq 
with $I_n(n')$ describing the impact of region $n'$ on region $n$. Here $\mb{\hat{x}}_n^{(s)}(t)$ and ${\mb{\hat{x}}_n}'^{(s)}(t)$ denote the predictions in region $n$ with and without the perturbation of $n'$ in the input space respectively, of a test sample $s$ at time step $t$.  

\begin{figure}
\bc
\includegraphics[width=0.9\textwidth]{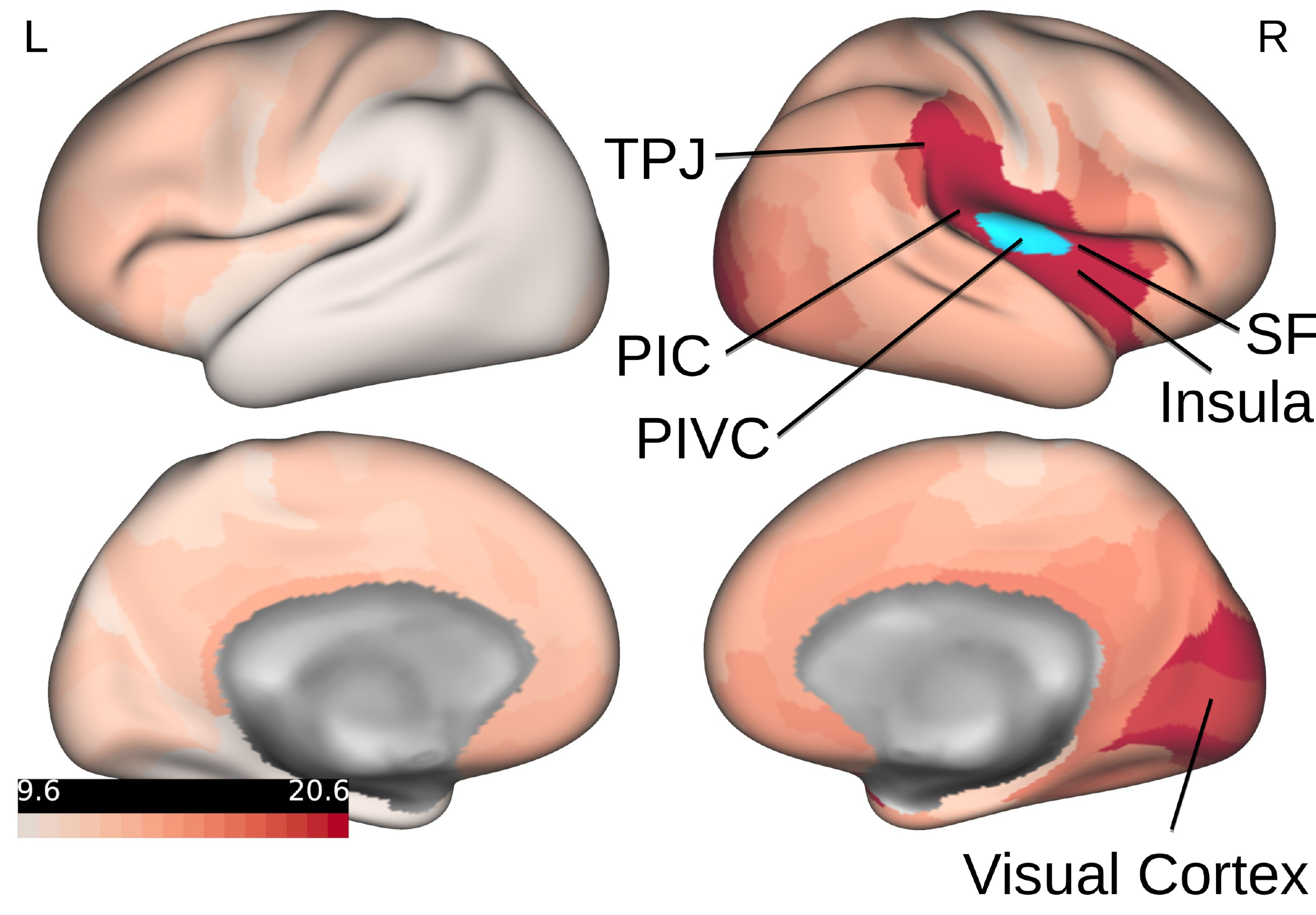}
\ec
\caption{The figure illustrates the influence of activity in PIVC on all other brain regions. The left side depicts the left hemisphere, while on the right side the right hemisphere is shown. The target region PIVC in the right hemisphere is marked in blue. The values of the influence measure $\mb{I}(n')$ were normalized between 0 and 100 and are encoded in red in this illustration. PIC = posterior insular cortex. PIVC = parieto-insular vestibular cortex. SF = Sylvian fissure and surrounding perisylvian cortex. TPJ = temporo-parietal junction. Note that causal relationships from right PIVC were primarily found in the ipsilateral hemisphere.}
\label{fig:ec_pivc}
\end{figure}

To visualize this measure of influence of $n'$ on each individual region $n$, values of $\mb{I}(n')$ can be projected onto the cortical surface. In the following we studied the impact of the parieto-insular vestibular cortex (PIVC) on all other brain regions. Here PIVC in the right hemisphere is characterized as a conjunction of ROIs R\_OP2-3 and R\_Ig, as defined by Glasser et al. \cite{Glasser2016}. Previous results show that this location coincides with the average location of PIVC across human subjects \cite{Lopez2012, Frank2018}. The perturbation was induced in R\_OP2-3 and R\_Ig simultaneously, and figure \ref{fig:ec_pivc} illustrates the strength of influence on all other regions (encoded in red) of the target region PIVC (marked in blue).

The results of this analysis show that PIVC exhibited strong causal connectivity with the Sylvian fissure, the perisylvian cortex and the insula. Similar connectivity patterns have been observed using diffusion weighted imaging \cite{Wirth2018, Indovina2020} and resting state functional connectivity \cite{Frank2020} in human subjects as well as in non-human primates using tracer techniques \cite{Guldin1998}. Several separate regions of the vestibular cortex are located within this Sylvian network, including the posterior insular cortex area (PIC), a region critical to the integration of visual and vestibular cues (for human subjects: \cite{Frank2014, Frank2016b}; for non-human primates the region is referred to as VPS: \cite{Guldin1998, Chen2011}). The information flow within this Sylvian network is not fully understood yet. Current theories assume that vestibular and visual cues about self motion are combined within PIVC and PIC and are then further processed to the temporoparietal junction (TPJ), a larger cortical region located at the junction of the temporal and parietal cortices, where visual-vestibular signals are integrated into a representation of the self in space \cite{Frank2018}. The results of the current analysis support this view by showing a causal relationship with the supramarginal gyrus, which is part of the TPJ. Further functional connectivity from PIVC was observed with the visual cortex. This result is interesting, since several studies have shown inhibitory interactions between the visual system and PIVC \cite{Wenzel1996, Brandt1998, Frank2016a, Frank2020}, such that PIVC is inhibited when visual cues are processed attentively and vice versa. These inhibitory interactions are assumed to be modulated in magnitude by attention networks located in the visual and parietal cortices (see \cite{Frank2020}).

%%%%%%%%%%%%%%%%%%%%%%%%%%%%%%%%%%
%%%%%%%%%%%%%%%%%%%%%%%%%%%%%%%%%%

\subsection{Model generalization.}

Often one problem is the availability of a sufficient amount of data, in order to fully train and take advantage of machine learning models with large parameter spaces. Especially in MRI studies it is usually time-consuming and costly to acquire such large data sets. To account for this limitation, the concept of transfer learning was proposed in machine learning \cite{Pan2010}. The basic idea behind transfer learning is that if only few data are available to learn a certain task, one can pretrain the model on a large-scale dataset of a similar task. In a next step, the feature representations learned on the large database can be used as an initialization for learning the desired target task. The goal is to transfer knowledge of one source domain to a target domain, by re-using the pretrained models weights. If the feature representation of the source domain is diverse enough, this can improve the model performance in comparison to starting the training without any prior knowledge, e.g. relying on a random initialization of the model weights \cite{Pan2010}.

To investigate, if transfer learning can also be suitable for our application, we studied the capabilities of the DCRNN to generalize across different datasets. Therefore we pretrained the DCRNN using the data provided by the HCP \cite{VanEssen2013}, as described in section \ref{sec:HCP_data}. The model was pretrained for $70$ epochs on in total 100 resting-state fMRI sessions, including the anatomical connectivity as reconstructed from DTI. Next we used a dataset collected with a \textit{Siemens Magnetom Prisma 3T} at the University of Regensburg (UR), including 10 resting-state fMRI sessions and corresponding structural imaging data. The acquisition parameters of this dataset are outlined in more detail in section \ref{sec:UR_data}. We fine tuned the DCRNN, pretrained on the HCP data, by training it for $70$ more epochs on the UR dataset, and initialized the second training with a lower learning rate of $0.001$. This pretrained model was compared to the DCRNN, only trained on the UR dataset, and with weight parameters initialized randomly with Xavier/Glorot initialization \cite{Glorot2010}.

\begin{figure}
\bc
\includegraphics[width=0.9\textwidth]{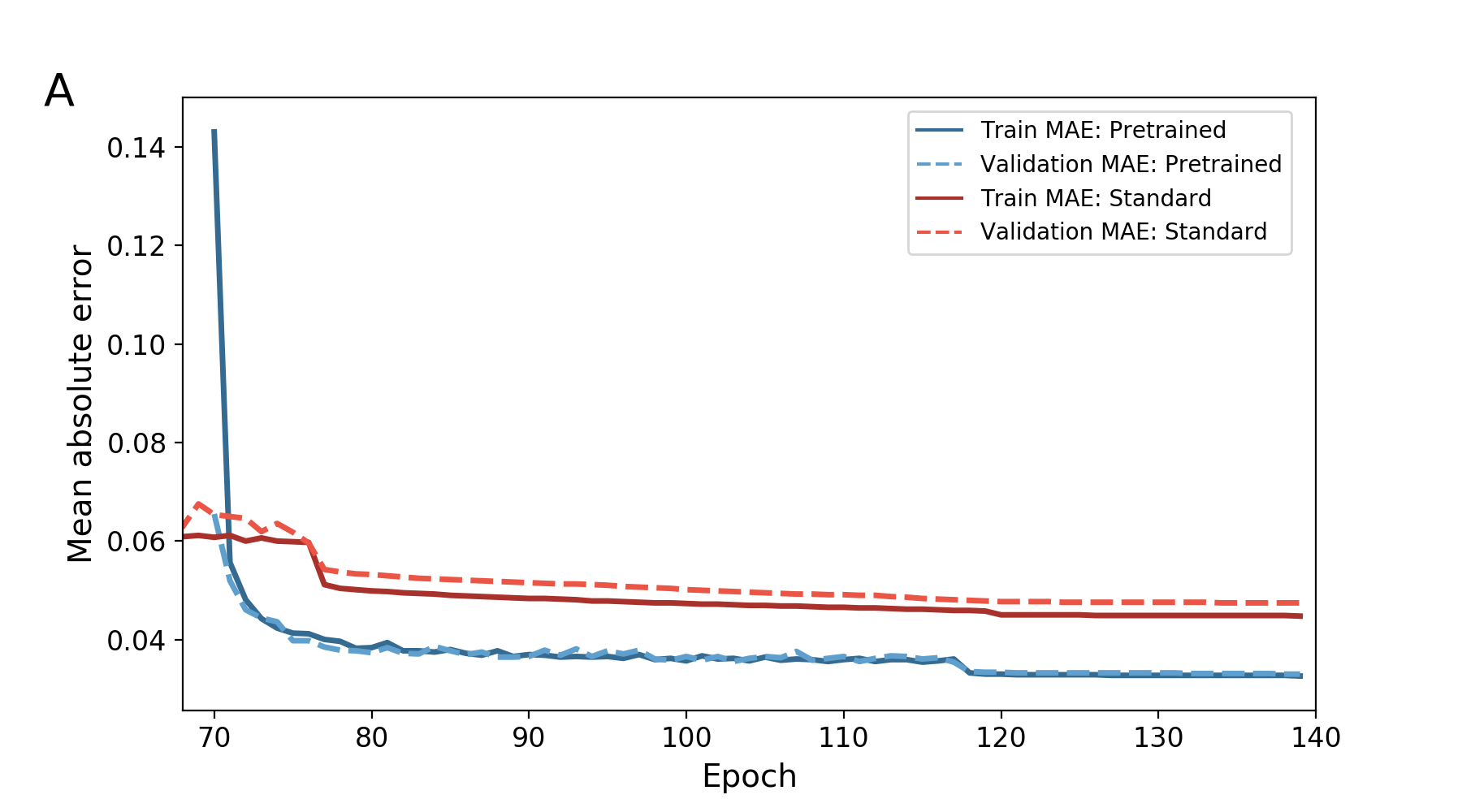}

\includegraphics[width=0.9\textwidth]{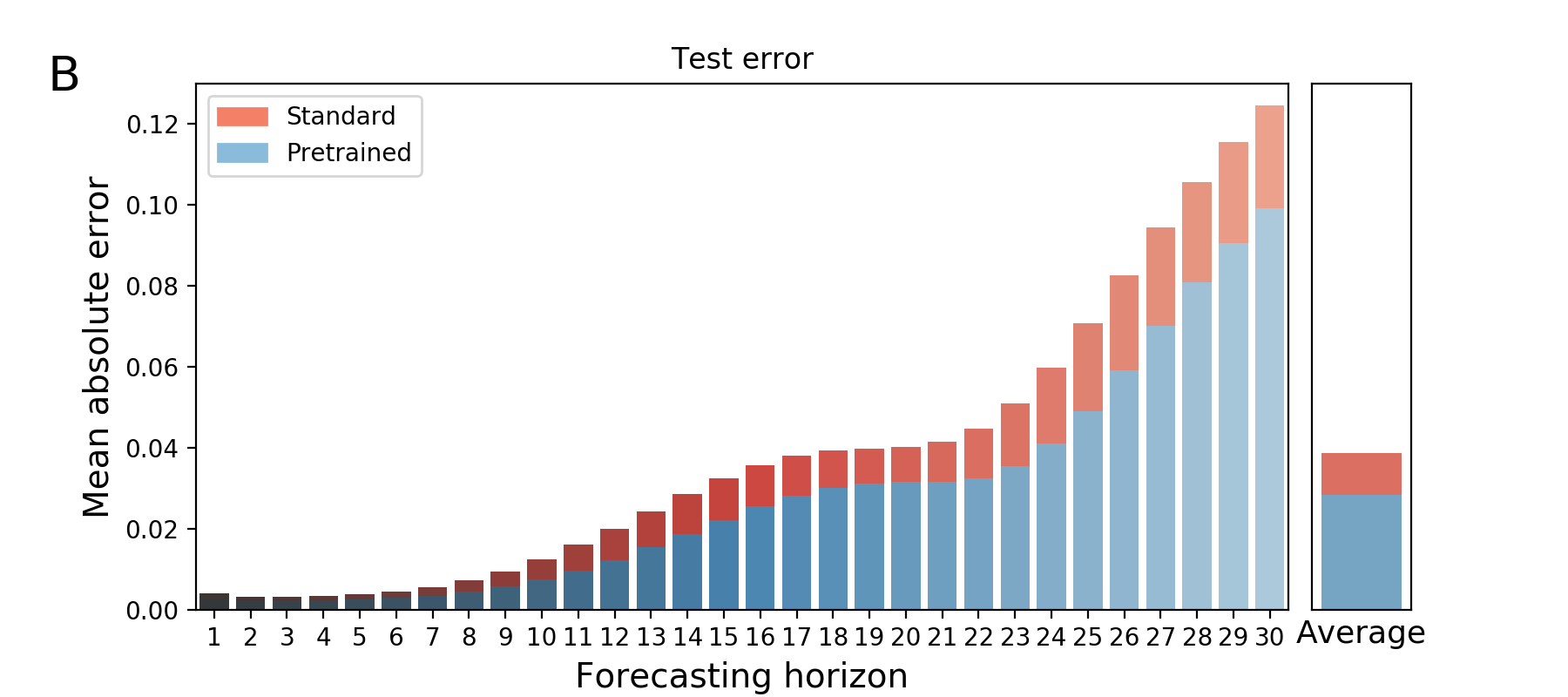}
\ec
\caption{This figure illustrates the performance difference between standard training and encompassing transfer learning. Figure (A) shows the validation and training MAE during learning from epoch 70 onwards, and the errors with and without pretraining are depicted in blue and red respectively. At the very beginning of fine tuning, the error of the pretrained model is relatively high, but decreases after the model adapts to the UR dataset. In figure (B) the final test MAE of both models is shown in dependence on the forecasting horizon.}
\label{fig:transfer_learning}
\end{figure}

The comparison between relying on standard training, and utilizing transfer learning is illustrated in figure \ref{fig:transfer_learning}. Figure \ref{fig:transfer_learning} (A) shows the training and validation error during learning when starting with a random initialization of the weights in red. This model was trained in total for $140$ epochs on the UR dataset only. In blue the training and validation error is depicted of the model, pretrained on the HCP dataset for $70$ epochs at first, and  fine tuned on the UR dataset for the subsequent $70$ epochs. Figure \ref{fig:transfer_learning} (A) illustrates that at the beginning, the training error on the UR data is relatively high, but as the pretrained model adapts to the new dataset the MAE becomes considerably smaller than without pretraining. In figure \ref{fig:transfer_learning} (B) the test MAE in dependence of the prediction horizon is depicted. The average test error could be reduced by $27 \%$ from $0.0388$ to $0.0284$ by encompassing transfer learning. Therewith the model performance on the small UR dataset, containing $10$ sessions a $7.3\min$, becomes comparable to the performance on the large HCP dataset with 100 sessions a $14.4\min$ with a $MAE=0.0279$.

\clearpage

%%%%%%%%%%%%%%%%%%%%%%%%%%%%%%%%%%
%%%%%%%%%%%%%%%%%%%%%%%%%%%%%%%%%%
%%%%%%%%%%%%%%%%%%%%%%%%%%%%%%%%%%

\section{Discussion}

We introduced a multi-modal framework for inferring causal relations in brain networks, based on a graph neural network architecture, uniting structural and functional information observed with DTI and fMRI. First this model provides a data-driven perspective on a fundamental question in neuroscience: How the function of the brain is related to its structure. Further by modeling dynamic interactions on the structural anatomical substrate, this framework accounts for non-linear spatio-temporal dependencies between segregated brain regions, allowing us to reconstruct a multi-modal measure of causal influence strength.

First, we evaluated the performance of the DCRNN by studying its capabilities to reproduce empirically observed neural activity patterns, and compared it to a VAR model, like typically used for the analysis of brain connectivity with Granger causality \cite{Granger1969, Seth2015}. We showed that the DCRNN can also capture temporal long-term dependencies in fMRI data, enabling it to make accurate predictions up to $30 \, TRs$ ($\approx 20s$) in the $0.04 - 0.07 Hz$ frequency range, what could reduced the overall test MAE considerably in comparison to a linear VAR. Note that results in \ref{sec:performance} demonstrate, that despite its simplicity, a VAR can make quite reliable predictions within the first $10 \, TRs$. Also its linearity allows for various possibilities for statistical inference of causal relations between different time courses, making it a practicable and fast tool for the estimation of Granger-causal connectivity \cite{Barnett2013}. But for the future it could be of interest to also consider non-linear and long-term relationships in neuroimaging data, in order to get a more complete picture of functional interactions between areas in the brain. The improved accuracy of the DCRNN reveals that it has better learned inherent characteristics of brain dynamics, and might be therefore more appropriate to characterize causal relations than simple linear models. We additionally verified the results by employing a more liberal bandpass filter with cutoff frequencies $0.02 - 0.09 Hz$ in supplement \hyperref[sec:supplement_1]{I}. By including more frequency components, the BOLD signal becomes more complex and is accordingly harder to predict. The same analysis has been carried out relying on a volumetric brain atlas \cite{Tzourio2002}, and using an alternative tractography method to reconstruct the structural connectivity in supplement \hyperref[sec:supplement_2]{II}. In all cases the difference between the VAR and DCRNN in the prediction performance is apparent, especially for larger horizons. Also the DCRNN does not require stationarity of time series data, therefore avoiding potentially distorting pre-pocessing steps in order to achieve the latter. Another aspect that improves the plausibility of the estimated causal relations between brain regions is the integration of structural information into the graph neural network model. As the propagation of neural signals is physically constrained by the layout of white matter connections, propagating information via graph convolutions along anatomically connected regions is in agreement with prior knowledge on the anatomy of the brain. 

The impact of this structural modeling was further investigated in section \ref{sec:impact_structure}. In the DCRNN the propagation of information is realized as a stationary diffusion process in the notion of a diffusion convolutions (DCs) \cite{Li2018}. Results show that diffusion steps of order $K=1$ already contribute most to the improvement of prediction accuracy, while higher order terms of $K=2,3$ only have a nominal further impact on the performance. The influence of structural modeling on the predictive performance provides additional insight into  the general structure-function relation in the brain, by pointing out, how much additional information about the functional activity in a certain region can be gained from the inclusion of structurally connected regions. By including filters up to order $K=3$, the predictions could be improved by $27\ \%$ in comparison to when information from anatomically connected regions has been neglected. Note that for each time step $t$ the DCRNN already applies multiple DCs to the multi-variate time series data, thereby inherently capturing the influence of higher order structural connections. Therefore low orders of diffusion walks $K \le 3$ seem to be already sufficient to account for indirect transitions. A good trade-off between computational load and model accuracy could be achieved with a maximum walk order of $K=2$, as the computational complexity linearly increases with $K$. Learning localized filters characterized by polynomial coefficients $\theta_k$ renders it possible to efficiently analyze large scale networks \cite{Defferrard2016}, what allowed us to conduct an analysis with $N=360$ regions simultaneously on a single GPU. So unlike classical DCM, this model can also be applied to study interactions across the whole brain, making it suitable for an exploratory analysis.   

The results demonstrated that propagating information across anatomical connections improves the model accuracy, pointing towards functional dependencies between different brain regions. In the spirit of explainable artificial intelligence (XAI), we proposed a method to reconstruct such dependencies, which the DCRNN has learned from the data in section \ref{sec:EC}. Inducing perturbations in the model's input space allowed us to study how the activity in a certain region influences other regions. This influence would quantify the importance of temporal information on the activity in a certain ROI for predicting the activity in other ROIs. Following the philosophy of Granger causality, this indicates a causal dependency between ROIs, thereby providing a measure of directed influence among each other. This kind of relation is referred to as \textit{directed functional connectivity} or \textit{causal connectivity}, as such information theoretic measures are dependent on causal mechanisms, but are not necessarily identical with them \cite{Seth2012, Bressler2010}, which distinguishes them from explicit model-based approaches like DCM for \textit{effective brain connectivity} \cite{Friston2013}. For our approach we used the more general notion of \textit{causal connectivity}, as we do not only incorporate functional data, but also structural information to describe such causal dependencies between different regions. To demonstrate an application of our proposed approach, we evaluated the influence of PIVC on other brain regions. This could reveal a causal relation of PIVC with brain regions in the Sylvian fissure, the perisylvian cortex and the insula, but also with the visual cortex. 

In a final step, we proposed an approach to improve the model performance on smaller data sets. We demonstrated that the concept of transfer learning \cite{Pan2010} finds also an application in our context of detecting intrinsic patterns in fMRI time-series and structural connectivity data. Features learned from the data of the HCP repository \cite{VanEssen2013} could be well transferred to a smaller dataset, acquired with a \textit{Siemens Magnetom Prisma 3T}. This made it possible to achieve an almost as good accuracy on a small dataset with 10 sessions (each 7.3 minutes in duration) as with a large dataset of 100 sessions (each 14.4 minutes in duration). The acquisition and preprocessing protocols of the two fMRI datasets were relatively comparable in our study, so in other cases with larger differences in the temporal resolutions of the data, downsampling one dataset could be necessary in order to better learn transferable feature representations.

Note that by integrating the structural information into the model, the functional interactions learned by this model also depend on the predefined anatomical layout. Therefore the quality of DTI data is additionally relevant for the results, but it is known that DTI has problems to accurately reconstruct long-range white matter tracks \cite{Thomas2014}. Also fMRI comes with its limitations for studying neural interactions, as the sampling rate is considerably lower than the underlying neural responses, and the neural activity is only indirectly measured based on the observed hemodynamic response \cite{Seth2015}. So the interpretation of the results should consider the informative content of the neuroimaging data used in this model. 

In conclusion we think that GNN architectures can provide an interesting novel approach to combine complex non-linear temporal and spatial patterns as observed in fMRI and DTI data. Currently GNNs already show very promising applications for classification tasks in MRI based on brain connectivity networks \cite{Ktena2018, Arslan2018, LiX2019, Kim2020}. In our study we showed that they can be also suitable to characterize the non-Euclidean spatial relationship of segregated brain regions when analyzing dynamic functional interactions on the structural network. Beyond the investigation of causal relations, this data-driven approach for brain dynamics could also be of interest for other applications. While many current approaches dealing with the structure-function relationship in the human brain focus on inferring the overall functional coherence patterns from their SC \cite{Becker2018, Surampudi2018, Saggio2016, Messe2014, Liang2017, Deligianni2016}, this framework allows us to directly relate temporally resolved activity profiles to their anatomical substrate. Further this whole-brain model could be of use for clinical applications, by studying dynamics in the diseased brain or modeling the impact of structural lesions \cite{Saenger2017, Alstott2009}. For future studies it could also be interesting to apply this multi-modal GNN model to other functional neuroimaging modalities like those obtained with electroencephalography (EEG) or magnetoencephalography (MEG) to investigate brain dynamics with greater temporal resolution.

%%%%%%%%%%%%%%%%%%%%%%%%%%%%%%%%%%
%%%%%%%%%%%%%%%%%%%%%%%%%%%%%%%%%%
%%%%%%%%%%%%%%%%%%%%%%%%%%%%%%%%%%

%\clearpage

\section{Methods} \label{sec:methods}

\subsection{DCRNN.}

In the context of neuroimaging, neural activity patterns can be interpreted as a graph structured spatio-temporal signal distribution. The nodes in this graph represent ROIs in a human brain, while the edges reflect the connection strengths between these ROIs in the anatomical neuronal network, which forms a structural scaffold for the flow of information. This connection strength is given by the axonal connection strength as determined from DTI measurements. The activity dynamics on such networks can be modeled by a random walk on a graph, where a diffusion convolution operation is invoked to capture the spatial dependencies \cite{Li2018, Defferrard2016}. A diffusion-convolution recurrent neural network (DCRNN) is designed to integrate diffusion convolution, a sequence-to-sequence architecture and a scheduled sampling technique \cite{Li2018}. The model, as it is applied in the current study, is described in detail below.

When considering voxel time series of brain activity maps, we collect all data into a data matrix $\mb{X} = (\mb{x}(1) \ldots \mb{x}(T)),$ with $ \mb{x}(t) \in \mbb{R}^N$. Given $N$ ROIs, taken from a brain atlas and each represented by a meta-voxel, and considering $T$ time points for each meta-voxel time series, which represents the activation time course of one of the ROIs, then we have

\beq 
\mb{X} = \left( \begin{array}{ccc}
                 x_{11} & \cdots & x_{1T} \\
                 \vdots & x_{nt} & \vdots \\
                 x_{N1} & \cdots & x_{NT}
                \end{array}
 \right)
\eeq
Note that the columns $\mb{x}(t) \in \mbb{R}^N$ of the data matrix describe the activation of all ROIs at any given time point $1 \le t \le T$, while its rows $\tilde{\mb{x}}_n(t), \; t=1, \ldots ,T$ represent the meta-voxel time course of every ROI $1 \le n \le N$.

Now consider a network of ROIs (brain areas, neuron pools) as an {\em undirected graph} $\mc{G} = (\mc{V},\mc{E},\mb{A}_w)$, where $\mc{V}, |\mc{V}| = N$ denotes a set of vertices (nodes), $\mc{E}$ represents a set of edges and $\mb{A}_w \in \mbb{R}^{N \times N}$ is a {\em weighted adjacency matrix}. The latter represents the structural connectivity of the nodes, i.e. the ROIs on the neuronal network, which are adjacent to each other and connected by an edge. Such undirected graphs can be deduced from diffusion tensor imaging (DTI) data, which also provide the edge weights $w_{nn'}$. The latter reflect the anatomical connection strengths between the connected vertices. Note that DTI alone cannot determine the direction of information flow, what makes it necessary to incorporate functional imaging data. 

The flow of activity observed on $\mc{G}$ is expressed as a time-dependent graph signal $\mb{x}(t) \in \mbb{R}^{N}$. It represents the feature of each ROI, which here is the  BOLD signal amplitude. Forecasting the flow of activity on $\mc{G}$ amounts to learning a function $h(...)$ that maps $T_p$ past graph signals to future $T_f$ graph signals:

\beq 
[\mb{x}(t-T_p+1), \ldots, \mb{x}(t); \mc{G}] \xrightarrow{h(...)} [\mb{x}(t+1), \ldots, \mb{x}(t+T_f)]
\eeq 

%%%%%%%%%%%%%%%%%%%%%%%%%%%%%%%%%%
%%%%%%%%%%%%%%%%%%%%%%%%%%%%%%%%%%

\subsubsection{Spatial dependencies.}

Information flow on $\mc{G}$  is considered a stochastic random walk process modeled by 
\bit 
\item 
a re-start probability $\alpha \in [0,1]$
\item 
a state transition matrix $\mb{T} = \mb{D}^{-1} \mb{A}_w = \left( \hat{\mb{w}}_1 \ldots \hat{\mb{w}}_N \right)$
\eit 

\nit Here we have with $\mb{w} \in \mbb{R}^N$ and $\hat{\mb{w}}_n = \left( \hat{w}_{1n} \ldots \hat{w}_{Nn}  \right)^T \; \forall \; n =1, \ldots ,N$

\beqa  
\mb{D} &=& diag(\mb{A}_w \mb{1}) 
\eeqa 

\nit where the $\hat{w} = w_{nn'} / \sum_{n'} w_{nn'}$ denote normalized edge strengths. Here state transitions are modeled as a diffusion process on a graph. Note that because the DTI cannot obtain directed graphs, its diffusion matrix is symmetric, i. e. $\mb{T} = \mb{T}^T$. Thus an eigen-decomposition exists according to 

\beq 
\mb{T} = \mb{U}\bs{\Lambda}\mb{U}^T.
\eeq 
Further the state transition matrix $\mb{T}$ is proportional to a normalized graph Laplacian 

\beq 
\mb{L}_{rw} = \mb{I} - \mb{T} = \mb{U} \left( \mb{I} - \bs{\Lambda} \right) \mb{U}^T
\eeq 
representing a random walk on the graph. Now consider the set of eigenvectors $\mb{U}$ of the diffusion Laplacian matrix as a set of basis vectors. Then the graph signal $\mb{x}_t \in \mbb{R}^N $ can be transformed to the conjugate domain and vice versa, hence we have \cite{Shuman2013}

\beqa 
\mb{x}_{\omega} &=& \mb{U}^T \mb{x}_{t} \\
\mb{x}_{t} &=& \mb{U} \mb{x}_{\omega}
\eeqa
Finally invoking the convolution theorem, the {\em graph convolution operator} $*_G$ can be defined as

\beq 
\mb{y}_{t} = \mb{x}_t \ast_G \mb{f}_{\theta} = \mb{U} \left( (\mb{U}^T \mb{f}_{\theta}) \odot (\mb{U}^T \mb{x}_t) \right) = \mb{U} \left( \bs{\theta}_{\omega} \odot \mb{x}_{\omega} \right) 
\eeq 
where $\mb{f}_{\theta}$ denotes a filter parametrized by $\theta$ and $\odot$ denotes the Hadamar product in the conjugate domain. The transformed vector $\mb{U}^T \mb{f}_{\theta} \equiv \bs{\theta}_{\omega} = (\theta_1(\omega), \ldots ,\theta_N(\omega))^T$ summarizes the filter parameters $\theta_n, n = 1, \ldots , N$ into a parameter vector in the conjugate frequency domain. If it is replaced by a diagonal feature matrix, i. e. $\bs{\theta}_{\omega} \rightarrow \bs{\Theta}_{\omega} = \diag (\theta_1(\omega) \ldots \theta_N(\omega))$, it represents a convolution kernel. Thus we have for the output signal

\beq 
\mb{y}_t = \mb{U} \bs{\Theta}_{\omega}\mb{x}_{\omega} = \mb{U} \bs{\Theta}_{\omega} \mb{U}^T \mb{x}_t 
\eeq 

Now expanding the filter kernel $\bs{\Theta}_{\omega}$ into a power series with respect to the eigenvalue matrix $\bs{\Lambda}$ of the transition matrix $\mb{T}´$, unfolding the bi-quadratic form into a sum of rank one outer product forms $\theta_n \mb{u}\mb{u}^T , n=1, \ldots , N$, which can be considered elementary filter kernel, and finally keeping only terms up to order $K$, we obtain

\beqa \label{eq:DC}
\mb{y}_t &=& \mb{U} \left[ \left( \sum_{k=0}^{K} \theta_k(\omega) \bs{\Lambda}^k  \right) \mb{U}^T \mb{x}_t \right] \nonumber \\
&=& \sum_{k=0}^{K} \theta_k(\omega) \mb{T}^k \mb{x}_t \nonumber \\
\eeqa  

\nit Note that this diffusion convolution operation includes the inverse diffusion process, represented by the transpose state transition matrix $\mb{T}^T$ as well, since DTI can only yield undirected graphs.  Thus, as has been shown by \cite{Li2018}, diffusion convolution is intimately related to spectral graph convolution (SGC) \cite{Defferrard2016}. More precisely, GDC is equivalent to SGC up to a similarity transformation \cite{Li2018}.

Considering a CNN architecture and using the diffusion convolution operation, the output of each of the $q \in \{1,\ldots,Q\}$ diffusion convolution layers (DCL) is then given as follows:

\beq 
\mb{h}_{q,t} = \sigma \left( \mb{y}_{q,t} \right) = \sigma \left( \sum_{k=0}^{K} \theta_{k,q} \mb{T}^k \mb{x}_t \right)
\eeq 

\nit Hereby $\mb{x}_t \in \mbb{R}^{N}$ denotes the input at time $t$, $\mb{h}_{q,t} \in \mbb{R}^{N}$ the corresponding output of the $q$-th convolution layer, $Q$ the number of filters employed, $\sigma(...) $ any suitable activation function, and $\theta_{q,k} \in \mbb{R}^{K+1}$ parameterizes the $q$-th convolutional kernel of order $k$. The DCL learns to represent graph structured data and can be trained with gradient descent based optimization techniques.

Note that this random walk on a graph represents a Markov process. At the limit $K \to \infty$ it converges to a stationary distribution $\mb{P} \in \mbb{R}^{N \times N}$, which for finite $K < \infty$ can be approximated by \cite{Teng2016}

\beq 
\mb{P} = \sum_{k=0}^{K} \alpha (1 - \alpha)^k  \mb{T}^k
\eeq 

\nit  The $i$-th row $\mb{P}_{i,*}$ of this matrix represents the likelihood of diffusion starting from ROI $v_i \in \mc{V}$, hence the proximity of any other ROI $v_j \in \mc{V}$ with respect to ROI $v_i$. 

%%%%%%%%%%%%%%%%%%%%%%%%%%%%%%%%%%
%%%%%%%%%%%%%%%%%%%%%%%%%%%%%%%%%%

\subsubsection{Temporal dependencies.}
\begin{figure}[!htb]
\bc
\includegraphics[width=0.8\textwidth]{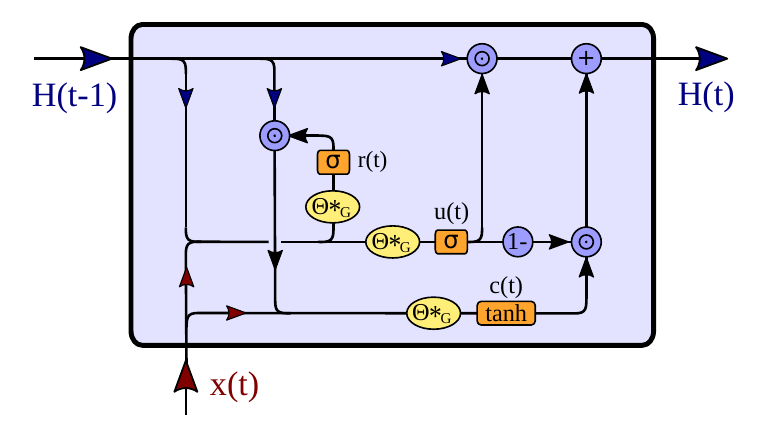}
\ec 
\caption{Overview of the processing steps of the DCGRU cell. The input $\mb{x}(t)$, as well as the previous hidden state $\mb{H}(t-1)$ are concatenated and passed to the reset gate $\mb{r}(t)$, as well as to the update gate $\mb{u}(t)$. The reset gate $\mb{r}(t)$ controls the proportion of $\mb{H}(t-1)$ which enters $\mb{c}(t)$, together with input $\mb{x}(t)$. Then the hidden state $\mb{H}(t-1)$ is updated by $\mb{c}(t)$, whereby the amount of new information is controlled by $\mb{u}(t)$.}
\label{fig:GRU}
\end{figure}

Given the graph convolution operation, temporal dynamics on the graph can be modeled using gated recurrent units (GRU) \cite{Chung2014}. The trick is to replace the matrix multiplications in GRU by diffusion convolutions $*_G$, as derived in equation \ref{eq:DC}. This leads to the diffusion convolutional gated recurrent unit (DCGRU) \cite{Li2018}

\beqa
\mb{r}(t) &=& \sigma \left( \bs{\Theta}_r *_G \left[ \mb{x}(t), \mb{H}(t-1)\right] + \mb{b}_r \right) \\
\mb{u}(t) &=& \sigma \left( \bs{\Theta}_u *_G \left[ \mb{x}(t), \mb{H}(t-1) \right] + \mb{b}_u \right) \\
\mb{c}(t) &=& \tanh \left( \bs{\Theta}_c *_G \left[ \mb{x}(t), (\mb{r}(t) \odot \mb{H}(t-1)) \right] + \mb{b}_c \right) \\
\mb{H}(t) &=& \mb{u}(t) \odot \mb{H}(t-1) + (1 - \mb{u}(t)) \odot \mb{c}(t)
\eeqa 
where $\mb{x}(t), \mb{H}(t)$ denote the input and output states of the GRU at time $t$ and $[\mb{x}(t),\mb{H}(t-1)]$ denotes their concatenation. Also $\mb{r}(t), \mb{u}(t)$ represent reset and update gates at time $t$, and $\mb{b}_r, \mb{b}_u, \mb{b}_c$, respectively, denote bias terms. Furthermore, $\bs{\Theta}_r, \bs{\Theta}_u, \bs{\Theta}_c$ denote the parameter sets of the corresponding filters. An illustration of a single DCGRU cell is provided in figure \ref{fig:GRU}.

Similar to GRUs, also DCGRUs can be employed to build layers of recurrent neural networks, which can be trained by backpropagation through time (BPTT) \cite{Werbos1990, Lillicrap2019}. If multiple step ahead forecasting is intended, a  sequence-to-sequence architecture can be used. In this architecture, both the encoder and the decoder are composed of DCGRU layers forming a diffusion convolution recurrent neural net (DCRNN) (see Fig. \ref{fig:DCRNN}). During training, a time series of past events is fed into the encoder and its final states form the input to the decoder. The latter then generates predictions, which can be compared to available ground truth observations. For later testing, such ground truth observations are replaced by predictions generated by the model itself. Given BOLD signal voxel activation time series, segments of an observed voxel time series are used to train a DCRNN to predict future activations. 

%%%%%%%%%%%%%%%%%%%%%%%%%%%%%%%%%%
%%%%%%%%%%%%%%%%%%%%%%%%%%%%%%%%%%

\subsubsection{Training the DCRNN.}

\begin{figure}[!htb]
\bc
\includegraphics[width=0.95\textwidth]{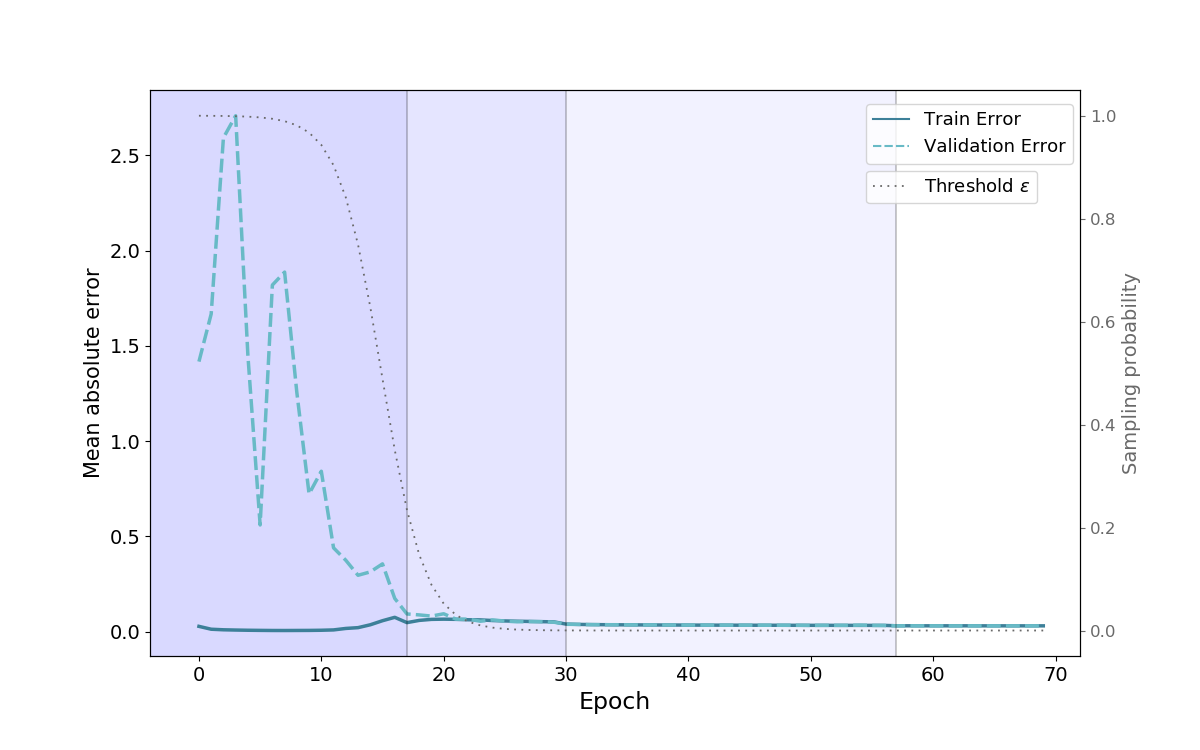}
\ec 
\caption{Illustration of the model performance during training. The figure shows the MAE during the training (solid blue line) and validation data set (dashed light blue line) in dependence of the number of epochs. The gray line illustrates the scheduled sampling probability $\epsilon_i$ over time. Vertical lines indicate when the learning rate was lowered by a factor of $0.1$. In the first few epochs the training error, due to the high schedule sampling probability $\epsilon_i$, is already quite low. During testing and validation the inputs for the decoder are always the models own predictions, what reflects the large discrepancy between training and validation error within the first epochs. When the sampling probability is subsequently decreased, the model also learns to successfully make long term forecasting.}
\label{fig:training}
\end{figure}

The network is trained by maximizing the likelihood of generating the target future time series using BPTT learning. Hence, DCRNN can capture spatio-temporal dependencies between time series. The DCRNN \cite{Li2018} was implemented using the \textit{TensorFlow} \cite{Abadi2016} library for machine learning, and computations were performed on an \textit{Nvidia Quadro K6000} GPU, running on a desktop PC with an \textit{Intel(R) Xeon(R) CPU E5-1620 v4} CPU under \textit{Linux Debian 9}. Scheduled sampling \cite{Bengio2015} is invoked during training to account for the fact that the distribution of input stimuli during testing might differ from the distribution of training stimuli. During scheduled sampling reference observations are fed to the model with probability $\epsilon_i$, while predictions released by the model are fed in with probability $1 - \epsilon_i$ at the $i$-th iteration. During supervised training, instances to be predicted are, of course, known. An inverse sigmoidal function determines the sampling probability decay:

\beq
	\epsilon_i = \frac{\tau}{1-exp(i/\tau)}
\eeq 
It was found to be sufficient to train the model for $70$ epochs, and the scheduled sampling parameter can be set to $\tau=5000$. As an objective function the mean absolute error (MAE) was used to describe the overall difference between true activity $\mb{x}(t)$ and predicted activity $\hat{\mb{x}}(t)$:
\beq
\mb{MAE}(\mb{x} , \hat{\mb{x}}) = \sum_{t=1}^{T_f} |\mb{x}(t) - \hat{\mb{x}}(t)|
\eeq
For this optimization problem, the ADAM algorithm \cite{Kingma2014} was employed, and the gradient was derived from mini-batches of $32$ samples. To further improve convergence, an annealing learning rate was used, initialized as $\eta = 0.1$, and decreased by a factor of $0.1$ at epochs $20$, $40$ and $60$, or if the validation error did not improve for more than $10$ epochs. Before lowering the learning rate, the weights with lowest validation error were restored, in order to avoid getting stuck in local optima. The encoder and decoder of the sequence-to-sequence architecture consist to two diffusion convolution GRU layers each, and the hidden state size is set to $Q=64$. The training performance is illustrated in figure \ref{fig:training}.

%%%%%%%%%%%%%%%%%%%%%%%%%%%%%%%%%%
%%%%%%%%%%%%%%%%%%%%%%%%%%%%%%%%%%

\subsection{Autoregressive models.} \label{sec:VAR}

As Granger causality \cite{Granger1969} is typically based on linear vector autoregressive (VAR) models for stochastic time series data, we evaluated a VAR as one baseline method. The idea of an autoregressive process (AR) is that a time series $x(t)$ can be described by a linear function of the first $T_p$ of its lagged values \cite{Luetkepohl2005}

\beq
	x(t) = \beta + \alpha_1 x(t-1) + \alpha_2 x(t-2) + \dots + \alpha_p x(t-T_p) + u(t) 
\eeq
with coefficients $ \alpha_1 \dots \alpha_p $, intercept $\beta$ and an error term $u(t)$. This expression can be extended to a multivariate VAR model with $N$ time series $\mb{x}(t) = [x_1(t), \dots , x_N(t)]$ as  \cite{Luetkepohl2005} 

\beq
	\mb{x}(t) = \mb{b} + \mb{A}_1 \mb{x}(t-1) + \mb{A}_2 \mb{x}(t-2) + \dots + \mb{A}_p \mb{x}(t-T_p) + \mb{u}(t) 
\eeq
where coefficients are stored in matrices $ \mb{A} \in \mbb{R}^{N \times N} $, and intercepts and errors are described by vectors $\mb{b} \in \mbb{R}^{N} $ and $\mb{u}(t)  \in \mbb{R}^{N}$. In the context of this study, time series $\mb{x}(t)$ reflect the BOLD signal of $N$ brain regions, measured at different times $t$.  

For the estimation of coefficients $\mb{A}$ and intercepts $\mb{b}$ various methods exist \cite{Barnett2013}, and in this study we rely on two different strategies. The first is based on a typical ordinary least squares (OLS) estimation \cite{Hamilton1994, Barnett2013} on individual subject sessions, implemented in the \textit{statsmodel} python package \cite{Seabold2010}. The first $80 \%$ of each fMRI session were used to fit the model to the data, and the subsequent $10 \%$ were used for validation and the last $10 \%$ were employed as a test set. To check for stationarity of the analyzed time series an augmented Dickey-Fuller test for unit roots was performed \cite{Hamilton1994, Mackinnon1994}, with a p-value of $p<0.01$.

Additionally, in order to render the comparison to the DCRNN more accurate, we implemented a gradient descend based optimization for a VAR model in \textit{TensorFlow} \cite{Abadi2016}, to verify that the differences in predictive performance can be related to the models, and not solely to the optimization strategies. In analogy to the DCRNN training, input-output samples of neural activities were generated from the data like described in section \ref{sec:data_descr}, which were used to minimize the MAE between the model's prediction $\mb{\hat{x}}(t)$ and groundtruths $\mb{x}(t)$. The convergence could be optimized by employing stochastic gradient descent (SGD) optimization with a batch size of $1$, using an annealing learning rate with a start value of $\eta = 0.005$. The VAR model was trained for $100$ epochs, and the learning rate was reduced by a factor of $0.1$ after epoch $70$ and $90$. A comparison of the error on the test set between the two different optimization strategies can be found in supplement \hyperref[sec:supplement_3]{III}. 

Best performance could be achieved employing a SGD based optimization in combination with a lag order of $P=30$. Note that with such a high lag order, around $9.7\%$ of the $N=360$ time courses do not fulfill the stationarity criteria of the augmented Dickey-Fuller test anymore ($p>0.01$). Yet the prediction accuracy could still be improved by including lags up to $P=30$, like shown in supplement \hyperref[sec:supplement_3]{III}. As the objective criterion of the evaluation was to assess the capabilities of replicating empirically observed neural activity patterns, we chose the VAR model with best accuracy for comparison with the DCRNN in section \ref{sec:performance}.

%%%%%%%%%%%%%%%%%%%%%%%%%%%%%%%%%%
%%%%%%%%%%%%%%%%%%%%%%%%%%%%%%%%%%

\subsection{Datasets.}

%%%%%%%%%%%%%%%%%%%%%%%%%%%%%%%%%%
%%%%%%%%%%%%%%%%%%%%%%%%%%%%%%%%%%

\subsubsection{HCP data.} \label{sec:HCP_data}

The first data set used in this study is provided by the HCP data repository \cite{Hodge2015, VanEssen2013}. The S1200 release includes data from subjects which participated in four resting state fMRI sessions, lasting $14.4\ min$ each and collecting $1200$ volumes per session. Customized \textit{Siemens Connectome Skyra} magnetic resonance imaging scanners with a field strength of $B_0 = 3 T$ were employed for data acquisition, using multi-band (factor 8) acceleration \cite{Moeller2010, Feinberg2010, Setsompop2012, Xu2012}. The data was collected by gradient-echo echo-planar imaging (EPI) sequences with a repetition time $TR = 720\ ms$ and an echo time $TE = 31.1\ ms$. The field of view was $FOV = 208\ mm \times 180\ mm$ and $N_s = 72$ slices with a thickness of $d_s = 2\ mm$ were obtained, containing voxels with a size of $2\ mm \times  2\ mm \times 2\ mm$. The preprocessed version, including motion-correction, structural preprocessing and ICA-FIX denoising was chosen \cite{Glasser2013, Jenkinson2002, Jenkinson2012, Fischl2012, Smith2013, Salimi2014, Griffanti2014}. Next a multi-model parcellation scheme was applied to divide the cortical gray matter hemisphere into 180 regions \cite{Glasser2016}, and the BOLD signal inside each region was averaged, to obtain the temporal activity evolution for each area.
For our study we found it appropriate to apply global signal regression, firstly because it showed to effectively reduce movement artifacts in HCP datasets \cite{Burgess2016}. 
%and additionally strengthens the relationship to anatomical connectivity \cite{Fox2009}. 
Also in this study of causal relations, the goal was to extract the additional information, which certain brain regions contain about the activity of other regions, whereby local interactions rather than global modulations were of interest for us. Those time coures were bandpass filtered, first performing the evaluations on a noise reduced narrowband in section \ref{sec:performance}, employing a filter with cutoff frequencies $0.04-0.07 Hz$ \cite{Glearean2012, Bruckner2009, Biswal1995, Achard2006}, and additionally implementing a more liberal bandpass filter with cutoff frequencies $0.02-0.09 Hz$, as displayed in supplement \hyperref[sec:supplement_1]{I}. 

Diffusion MRI data was collected in 6 runs, whereby approximately 90 directions were sampled during each run, employing three shells of $b=1000, 2000,$ and $3000 \ s/mm^2$, including 6 $b=0$ images \cite{Sotiropoulos2013b}. A Spin-echo EPI sequence was employed with repetition time $TR = 5520\ ms$, echo time $ TE = 89.5\ ms$, and multi band factor 3. In total $N_s = 111$ slices were obtained, with field of view $FOV =  210\ mm \times 180\ mm$ and an isotropic voxel size of $1.25\ mm \times  1.25\ mm \times 1.25\ mm$. The preprocessing included intensity normalization across runs, EPI distortion correction, eddy-current corrections, removing motion artifacts, and gradient non-linearity corrections \cite{Glasser2013, Sotiropoulos2013, Andersson2003, Andersson2015, Andersson2015b}. To obtain the structural connectivity strengths between regions defined by Glasser et al. \cite{Glasser2016}, the \textit{MRtrix3} software package was employed \cite{Tournier2019}. Briefly
multi-shell multi-tissue constrained spherical deconvolution \cite{Jeurissen2014} was used to obtain the response functions for fiber orientation distribution estimation \cite{Tournier2004, Tournier2007}. Furher 10 million streamlines were created with anatomical constrained tractography \cite{SmithR2012} and spherical-deconvolution informed filtering was applied \cite{SmithR2013}, reducing the number of streamlines to 1 million\footnote{A detailed description of the structural connectome generation can be found in: \textit{Basic and Advanced Tractography with MRtrix for All Neurophiles}: \url{https://osf.io/fkyht/}}. The group structural connectome was computed as an average across the first 10 subjects, as the variance in the structural connectivity strength is relatively low across subjects \cite{Zimmermann2019}, while probabilistic tracktography methods are relatively computationally demanding.

%%%%%%%%%%%%%%%%%%%%%%%%%%%%%%%%%%
%%%%%%%%%%%%%%%%%%%%%%%%%%%%%%%%%%

\subsubsection{UR data.} \label{sec:UR_data}

The second dataset was acquired with a \textit{Siemens Magnetom Prisma} with field strength $B_0 = 3 \ T$ at the University of Regensburg (UR). The data of 10 different subjects were used, whereby resting state fMRI data were collected during a scanning time of $7.3\ min$. All subjects provided written informed consent and the study was approved by the local ethics committee of the University of Regensburg. An EPI sequence was employed using multi-band (factor 8) acceleration, sampling 600 volumetric images per run with a repetition time of $TR= 730 \ ms$ and an echo time of $TE=31\ ms$. The field of view was $FOV = 208 \ mm \times 208 \ mm$ and $N_s = 72$ slices with thickness of $d_s = 2 \ mm $ were collected, containing voxels with a size of $2 \ mm \times 2 \ mm \times 2 \ mm $. For preprocessing the HCP pipeline (version 4.0.0) was employed, as described by Glasser et al. \cite{Glasser2013}. To achieve good correspondence between the two datasets, the further preprocessing was also performed like outlined in section \ref{sec:HCP_data}. The fMRI time courses were averaged within each brain region of the multi-modal parcellation scheme \cite{Glasser2016}, and again global signal regression was applied. Finally those time courses were bandpass filtered within the noise reduced range of $0.04 - 0.07\ Hz$.

To reconstruct the anatomical connectivity, diffusion MRI data was collected in 4 runs, sampling approximately 90 directions, employing two shells with $b = 1500$ and $ 3000 \ s/mm^2$, and also including 7 $b=0$ images. The repetition time of the Spin-echo EPI sequence was $TR=3222 \ ms$ with an echo time $TE=89,2 \ ms$, employing a multi-band (factor 4) acceleration. Overall $N_s=92$ slices were collected, with a field of view $FOV = 210 \ mm \times 210 \ mm$, containing voxels with a size of $ 1.5 \ mm \times 1.5 \ mm \times 1.5 \ mm$. Preprocessing of the diffusion MRI data was based on the HCP guidelines \cite{Glasser2013}, and finally the anatomical connectivity matrices were obtained like in section \ref{sec:HCP_data} using constrained spherical deconvolution as provided in the \textit{MRtrix} package \cite{Tournier2019}. The group structural connectivity was computed as an average over the 10 subjects.    

%%%%%%%%%%%%%%%%%%%%%%%%%%%%%%%%%%%
%%%%%%%%%%%%%%%%%%%%%%%%%%%%%%%%%%%

\section*{Acknowledgement}

The work was supported by the DFG projects GR988/25-1 granted to MWG and WU392/9-1 granted to SWü, the Harris Foundation of Dartmouth College granted to MWG, and the Hanns Seidel Foundation granted to G-IH. Data were provided in part by the Human Connectome Project, WU-Minn Consortium (Principal Investigators: David Van Essen and Kamil Ugurbil; 1U54MH091657) funded by the 16 NIH Institutes and Centers that support the NIH Blueprint for Neuroscience Research; and by the McDonnell Center for Systems Neuroscience at Washington University.

%%%%%%%%%%%%%%%%%%%%%%%%%%%%%%%%%%%
%%%%%%%%%%%%%%%%%%%%%%%%%%%%%%%%%%%

\bibliographystyle{abbrv}
\bibliography{GCN}

\begin{thebibliography}{100}

\bibitem{Abadi2016}
M.~Abadi, P.~Barham, J.~Chen, Z.~Chen, A.~Davis, J.~Dean, M.~Devin,
  S.~Ghemawat, G.~Irving, M.~Isard, et~al.
\newblock Tensorflow: A system for large-scale machine learning.
\newblock In {\em 12th {USENIX} Symposium on Operating Systems Design and
  Implementation ({OSDI} 16)}, pages 265--283, 2016.

\bibitem{Abdelnour2014}
F.~Abdelnour, H.~U. Voss, and A.~Raj.
\newblock Network diffusion accurately models the relationship between
  structural and functional brain connectivity networks.
\newblock {\em NeuroImage}, 90:335--347, 2014.

\bibitem{Achard2006}
S.~Achard, R.~Salvador, B.~Whitcher, J.~Suckling, and E.~Bullmore.
\newblock A resilient, low-frequency, small-world human brain functional
  network with highly connected association cortical hubs.
\newblock {\em J Neurosci}, 3:e17, 2006.

\bibitem{Alstott2009}
J.~Alstott, M.~Breakspear, P.~Hagmann, L.~Cammoun, and O.~Sporns.
\newblock Modeling the impact of lesions in the human brain.
\newblock {\em PLoS computational biology}, 5:e1000408, 2009.

\bibitem{Amico2018}
E.~Amico and J.~Go{\~n}i.
\newblock Mapping hybrid functional-structural connectivity traits in the human
  connectome.
\newblock {\em Network Neuroscience}, 2:306--322, 2018.

\bibitem{Andersson2003}
J.~Andersson, S.~Skare, and J.~Ashburner.
\newblock How to correct susceptibility distortions in spin-echo echo-planar
  images: Application to diffusion tensor imaging.
\newblock {\em NeuroImage}, 20:870--88, 2003.

\bibitem{Andersson2015b}
J.~Andersson and S.~Sotiropoulos.
\newblock An integrated approach to correction for off-resonance effects and
  subject movement in diffusion {MR} imaging.
\newblock {\em NeuroImage}, 125:1063--1078, 2015.

\bibitem{Andersson2015}
J.~Andersson and S.~Sotiropoulos.
\newblock Non-parametric representation and prediction of single- and
  multi-shell diffusion-weighted {MRI} data using gaussian processes.
\newblock {\em NeuroImage}, 122:166--76, 2015.

\bibitem{Arslan2018}
S.~Arslan, S.~I. Ktena, B.~Glocker, and D.~Rueckert.
\newblock Graph saliency maps through spectral convolutional networks:
  Application to sex classification with brain connectivity.
\newblock In {\em GRAIL/Beyond-MIC@MICCAI}, 2018.

\bibitem{Barnett2013}
L.~Barnett and A.~Seth.
\newblock The {MVGC} multivariate granger causality toolbox: A new approach to
  granger-causal inference.
\newblock {\em Journal of neuroscience methods}, 223:50--68, 2013.

\bibitem{Becker2018}
C.~Becker, S.~Pequito, G.~Pappas, M.~Miller, S.~T.~Grafton, D.~S.~Bassett, and
  V.~Preciado.
\newblock Spectral mapping of brain functional connectivity from diffusion
  imaging.
\newblock {\em Scientific Reports}, 8, 12 2018.

\bibitem{Behrens2007}
T.~Behrens, H.~Berg, S.~Jbabdi, M.~Rushworth, and M.~Woolrich.
\newblock Probabilistic diffusion tractography with multiple fibre
  orientations: What can we gain?
\newblock {\em NeuroImage}, 34:144--55, 2007.

\bibitem{Bengio2015}
S.~Bengio, O.~Vinyals, N.~Jaitly, and N.~Shazeer.
\newblock Scheduled sampling for sequence prediction with recurrent neural
  networks.
\newblock In {\em NIPS}, pages 1171--1179, 2015.

\bibitem{Bettinardi2018}
R.~G. Bettinardi, G.~Deco, V.~M. Karlaftis, T.~J.~V. Hartevelt, H.~M.
  Fernandes, Z.~Kourtzi, M.~L. Kringelbach, and G.~Zamora-L\'opez.
\newblock How structure sculpts function: unveiling the contribution of
  anatomical connectivity to the brain’s spontaneous correlation structure.
\newblock {\em Chaos: An Interdisciplinary Journal of Nonlinear Science}, 27,
  2018.

\bibitem{Biswal1995}
B.~B. Biswal, F.~Z. Yetkin, V.~Haughton, and J.~S. Hyde.
\newblock Functional connectivity in the motor cortex of resting human brain
  using echo-planar {MRI}.
\newblock {\em Magnetic resonance in medicine}, 34 4:537--41, 1995.

\bibitem{Brandt1998}
T.~Brandt, P.~Bartenstein, A.~Janek, and M.~Dieterich.
\newblock Reciprocal inhibitory visual-vestibular interaction. visual motion
  stimulation deactivates the parieto-insular vestibular cortex.
\newblock {\em Brain : a journal of neurology}, 121:1749--58, 1998.

\bibitem{Bressler2010}
S.~Bressler and A.~Seth.
\newblock Wiener–granger causality: A well established methodology.
\newblock {\em NeuroImage}, 58:323--9, 2010.

\bibitem{Bruna2014}
J.~Bruna, W.~Zaremba, A.~Szlam, and Y.~Lecun.
\newblock Spectral networks and locally connected networks on graphs.
\newblock In {\em International Conference on Learning Representations
  ({ICLR}2014)}, CBLS, 2014.

\bibitem{Bruckner2009}
R.~Buckner, J.~Sepulcre, T.~Talukdar, F.~Krienen, H.~Liu, T.~Hedden,
  J.~Andrews-Hanna, R.~Sperling, and K.~Johnson.
\newblock Cortical hubs revealed by intrinsic functional connectivity: Mapping,
  assessment of stability, and relation to alzheimer's disease.
\newblock {\em The Journal of neuroscience : the official journal of the
  Society for Neuroscience}, 29:1860--73, 2009.

\bibitem{Burgess2016}
G.~Burgess, S.~Kandala, D.~Nolan, T.~Laumann, J.~Power, B.~Adeyemo, M.~Harms,
  S.~Petersen, and D.~Barch.
\newblock Evaluation of denoising strategies to address motion-correlated
  artifact in resting state {fMRI} data from the human connectome project.
\newblock {\em Brain Connectivity}, 6, 2016.

\bibitem{Chen2011}
A.~Chen, G.~Deangelis, and D.~Angelaki.
\newblock Convergence of vestibular and visual self-motion signals in an area
  of the posterior sylvian fissure.
\newblock {\em The Journal of neuroscience : the official journal of the
  Society for Neuroscience}, 31:11617--27, 2011.

\bibitem{Chu2018}
S.~Chu, K.~Parhi, and C.~Lenglet.
\newblock Function-specific and enhanced brain structural connectivity mapping
  via joint modeling of diffusion and functional {MRI}.
\newblock {\em Scientific Reports}, 8, 2018.

\bibitem{Chung2014}
J.~Chung, C.~Gulcehre, K.~Cho, and Y.~Bengio.
\newblock Empirical evaluation of gated recurrent neural networks on sequence
  modeling, 2014.

\bibitem{Daunizeau2009}
J.~Daunizeau, O.~David, and K.~Stephan.
\newblock Dynamic causal modelling: A critical review of the biophysical and
  statistical foundations.
\newblock {\em NeuroImage}, 58:312--22, 2009.

\bibitem{Deco2017}
G.~Deco, M.~L. Kringelbach, V.~K. Jirsa, and P.~Ritter.
\newblock The dynamics of resting fluctuations in the brain: metastability and
  its dynamical cortical core.
\newblock {\em Scientific Reports}, 7, 2017.

\bibitem{Deco2012}
G.~Deco, M.~Senden, and V.~Jirsa.
\newblock How anatomy shapes dynamics: a semi-analytical study of the brain at
  rest by a simple spin model.
\newblock {\em Frontiers in computational neuroscience}, 6:68, 2012.

\bibitem{Defferrard2016}
M.~Defferrard, X.~Bresson, and P.~Vandergheynst.
\newblock Convolutional neural networks on graphs with fast localized spectral
  filtering.
\newblock In {\em NIPS}, pages 3837--3845, 2016.

\bibitem{Deligianni2016}
F.~Deligianni, D.~Carmichael, G.~H~Zhang, C.~Clark, and J.~Clayden.
\newblock Noddi and tensor-based microstructural indices as predictors of
  functional connectivity.
\newblock {\em PloS one}, 11:e0153404, 04 2016.

\bibitem{Feinberg2010}
D.~Feinberg, S.~Moeller, S.~M~Smith, E.~Auerbach, S.~Ramanna, M.~Günther,
  M.~F~Glasser, K.~Miller, K.~Ugurbil, and E.~Yacoub.
\newblock Multiplexed echo planar imaging for sub-second whole brain {FMRI} and
  fast diffusion imaging.
\newblock {\em PloS one}, 5:e15710, 2010.

\bibitem{Fischl2012}
B.~Fischl.
\newblock Freesurfer.
\newblock {\em NeuroImage}, 62(2):774 -- 781, 2012.

\bibitem{Frank2014}
S.~Frank, O.~Baumann, J.~Mattingley, and M.~Greenlee.
\newblock Vestibular and visual responses in human posterior insular cortex.
\newblock {\em Journal of neurophysiology}, 112:2481--2491, 2014.

\bibitem{Frank2018}
S.~Frank and M.~Greenlee.
\newblock The parieto-insular vestibular cortex in humans: More than a single
  area?
\newblock {\em Journal of Neurophysiology}, 120:1438--1450, 2018.

\bibitem{Frank2016b}
S.~Frank, A.~Wirth, and M.~Greenlee.
\newblock Visual-vestibular processing in the human sylvian fissure.
\newblock {\em Journal of neurophysiology}, 116:263--271, 2016.

\bibitem{Frank2020}
S.~M. Frank, M.~Pawellek, L.~D. Forster, B.~Langguth, M.~Schecklmann, and M.~W.
  Greenlee.
\newblock Attention networks in the parietooccipital cortex modulate activity
  of the human vestibular cortex during attentive visual processing.
\newblock {\em The Journal of Neuroscience}, 40:1110 -- 1119, 2020.

\bibitem{Frank2016a}
S.~M. Frank, L.~Sun, L.~Forster, P.~U. Tse, and M.~W. Greenlee.
\newblock Cross-modal attention effects in the vestibular cortex during
  attentive tracking of moving objects.
\newblock {\em The Journal of Neuroscience}, 36:12720 -- 12728, 2016.

\bibitem{Friston2013}
K.~Friston, R.~Moran, and A.~K. Seth.
\newblock Analysing connectivity with granger causality and dynamic causal
  modelling.
\newblock {\em Curr Opin Neurobiol}, 23:172--178, 2013.

\bibitem{Friston2003}
K.~J. Friston, L.~Harrison, and W.~Penny.
\newblock Dynamic causal modelling.
\newblock {\em Neuroimage}, 19:1273--1302, 2003.

\bibitem{Fukushima1987}
K.~Fukushima.
\newblock A neural network model for the mechanism of selective attention in
  visual pattern recognition.
\newblock {\em Systems and Computers in Japan}, 18:102 -- 113, 1987.

\bibitem{Glasser2016}
M.~Glasser, T.~Coalson, E.~Robinson, C.~Hacker, J.~Harwell, E.~Yacoub,
  K.~Ugurbil, J.~Andersson, C.~Beckmann, M.~Jenkinson, S.~Smith, and
  D.~Van~Essen.
\newblock A multi-modal parcellation of human cerebral cortex.
\newblock {\em Nature}, 536, 2016.

\bibitem{Glasser2013}
M.~Glasser, S.~Sotiropoulos, J.~Wilson, T.~Coalson, B.~Fischl, J.~Andersson,
  J.~Xu, S.~Jbabdi, M.~Webster, J.~Polimeni, V.~DC, and M.~Jenkinson.
\newblock The minimal preprocessing pipelines for the human connectome project.
\newblock {\em NeuroImage}, 80, 2013.

\bibitem{Glearean2012}
E.~Glerean, J.~Salmi, J.~Lahnakoski, I.~Jääskeläinen, and M.~Sams.
\newblock Functional magnetic resonance imaging phase synchronization as a
  measure of dynamic functional connectivity.
\newblock {\em Brain connectivity}, 2:91--101, 2012.

\bibitem{Glorot2010}
X.~Glorot and Y.~Bengio.
\newblock Understanding the difficulty of training deep feedforward neural
  networks.
\newblock {\em Journal of Machine Learning Research - Proceedings Track},
  9:249--256, 2010.

\bibitem{Granger1969}
C.~W.~J. Granger.
\newblock Investigating causal relations by econometric models and
  cross-spectral methods.
\newblock {\em Econometrica}, 37:424--438, 1969.

\bibitem{Griffanti2014}
L.~Griffanti, G.~Salimi-Khorshidi, C.~F. Beckmann, E.~J. Auerbach, G.~Douaud,
  C.~E. Sexton, E.~Zsoldos, K.~P. Ebmeier, N.~Filippini, C.~E. Mackay,
  S.~Moeller, J.~Xu, E.~Yacoub, G.~Baselli, K.~Ugurbil, K.~L. Miller, and S.~M.
  Smith.
\newblock {ICA-based artefact removal and accelerated fMRI acquisition for
  improved resting state network imaging}.
\newblock {\em NeuroImage}, 95:232 -- 247, 2014.

\bibitem{Hamilton1994}
J.~Hamilton.
\newblock {\em Time Series Analysis}.
\newblock Princeton University Press, Princeton, NJ., 1994.

\bibitem{Hermundstad2013}
A.~M. Hermundstad, D.~S. Bassett, K.~S. Brown, E.~M. Aminoff, D.~Clewett,
  S.~Freeman, A.~Frithsen, A.~Johnson, C.~M. Tipper, M.~B. Miller, S.~T.
  Grafton, and J.~M. Carlson.
\newblock Structural foundations of resting-state and task-based functional
  connectivity in the human brain.
\newblock {\em Proceedings of the National Academy of Sciences of the United
  States of America}, 110 15:6169--74, 2013.

\bibitem{Hodge2015}
M.~Hodge, W.~Horton, T.~Brown, R.~Herrick, T.~Olsen, M.~Hileman, M.~McKay,
  K.~Archie, E.~Cler, M.~Harms, G.~Burgess, M.~Glasser, J.~Elam, S.~Curtiss,
  D.~Barch, R.~Oostenveld, L.~Larson-Prior, K.~Ugurbil, D.~Van~Essen, and
  D.~Marcus.
\newblock {ConnectomeDB} – sharing human brain connectivity data.
\newblock {\em NeuroImage}, 124, 2015.

\bibitem{Honey2009}
C.~J. Honey, O.~Sporns, L.~Cammoun, X.~Gigandet, J.~P. Thiran, R.~Meuli, and
  P.~Hagmann.
\newblock Predicting human resting-state functional connectivity from
  structural connectivity.
\newblock {\em Proceedings of the National Academy of Sciences of the United
  States of America}, 106 6:2035--40, 2009.

\bibitem{Indovina2020}
I.~Indovina, G.~Bosco, R.~Riccelli, V.~Maffei, F.~Lacquaniti, L.~Passamonti,
  and N.~Toschi.
\newblock Structural connectome and connectivity lateralization of the
  multimodal vestibular cortical network.
\newblock {\em NeuroImage}, 222:117247, 2020.

\bibitem{Jbabdi2012}
S.~Jbabdi, S.~Sotiropoulos, A.~Savio, M.~Graña, and T.~Behrens.
\newblock Model-based analysis of multishell diffusion {MR} data for
  tractography: How to get over fitting problems.
\newblock {\em Magnetic resonance in medicine : official journal of the Society
  of Magnetic Resonance in Medicine / Society of Magnetic Resonance in
  Medicine}, 68, 2012.

\bibitem{Jenkinson2002}
M.~Jenkinson, P.~Bannister, M.~Brady, and S.~Smith.
\newblock Improved optimization for the robust and accurate linear registration
  and motion correction of brain images.
\newblock {\em NeuroImage}, 17:825 -- 841, 2002.

\bibitem{Jenkinson2012}
M.~Jenkinson, C.~F. Beckmann, T.~E. Behrens, M.~W. Woolrich, and S.~M. Smith.
\newblock {FSL}.
\newblock {\em NeuroImage}, 62(2):782 -- 790, 2012.

\bibitem{Jeurissen2014}
B.~Jeurissen, J.-D. Tournier, T.~Dhollander, A.~Connelly, and J.~Sijbers.
\newblock Multi-tissue constrained spherical deconvolution for improved
  analysis of multi-shell diffusion {MRI} data.
\newblock {\em NeuroImage}, 103:411--426, 2014.

\bibitem{Kiefer1952}
J.~Kiefer and J.~Wolfowitz.
\newblock Stochastic estimation of the maximum of a regression function.
\newblock volume~23, pages 462--466, 1952.

\bibitem{Kim2020}
B.-H. Kim and J.~C. Ye.
\newblock Understanding graph isomorphism network for rs-{fMRI} functional
  connectivity analysis.
\newblock {\em Frontiers in Neuroscience}, 14:630, 2020.

\bibitem{Kingma2014}
D.~Kingma and J.~Ba.
\newblock Adam: A method for stochastic optimization.
\newblock 2014.

\bibitem{Ktena2018}
S.~I. Ktena, S.~Parisot, E.~Ferrante, M.~Rajchl, M.~C.~H. Lee, B.~Glocker, and
  D.~Rueckert.
\newblock Metric learning with spectral graph convolutions on brain
  connectivity networks.
\newblock {\em NeuroImage}, 169:431--442, 2018.

\bibitem{Lang2012}
E.~Lang, A.~Tomé, I.~Keck, J.~Gorriz, and C.~Puntonet.
\newblock Brain connectivity analysis: A short survey.
\newblock {\em Computational intelligence and neuroscience}, 2012.

\bibitem{Lecun2015}
Y.~LeCun, Y.~Bengio, and G.~Hinton.
\newblock Deep learning.
\newblock {\em Nature}, 521:436--44, 2015.

\bibitem{LiX2019}
X.~Li, N.~Dvornek, Y.~Zhou, J.~Zhuang, P.~Ventola, and J.~Duncan.
\newblock Graph neural network for interpreting task-fmri biomarkers.
\newblock pages 485--493, 2019.

\bibitem{Li2018}
Y.~Li, R.~Yu, C.~Shahabi, and Y.~Liu.
\newblock Diffusion convolutional recurrent neural network: Data-driven traffic
  forecasting, 2018.

\bibitem{Liang2017}
H.~Liang and H.~Wang.
\newblock Structure-{F}unction {N}etwork {M}apping and its {A}ssessment via
  {P}ersistent {H}omology.
\newblock {\em PLoS Computational Biology}, 2017.

\bibitem{Lillicrap2019}
T.~P. Lillicrap and A.~Santoro.
\newblock Backpropagation through time and the brain.
\newblock {\em Current Opinion in Neurobiology}, 55:82--89, 2019.

\bibitem{Lopez2012}
C.~Lopez, O.~Blanke, and F.~Mast.
\newblock The human vestibular cortex revealed by coordinate-based activation
  likelihood estimation meta-analysis.
\newblock {\em Neuroscience}, 212:159--79, 2012.

\bibitem{Luetkepohl2005}
H.~Luetkepohl.
\newblock {\em The New Introduction to Multiple Time Series Analysis}.
\newblock Springer, 2005.

\bibitem{Saggio2016}
M.~Luisa~Saggio, P.~Ritter, and V.~K~Jirsa.
\newblock Analytical operations relate structural and functional connectivity
  in the brain.
\newblock {\em PloS one}, 11:e0157292, 08 2016.

\bibitem{Mackinnon1994}
J.~Mackinnon.
\newblock Approximate asymptotic distribution functions for unit-root and
  cointegration tests.
\newblock {\em Journal of Business and Economic Statistics}, 12:167--76, 1994.

\bibitem{Messe2014}
A.~Mess\'e, D.~Rudrauf, H.~Benali, and G.~Marrelec.
\newblock Relating {S}tructure and {F}unction in the {H}uman {B}rain:
  {R}elative {C}ontributions of {A}natomy, {S}tationary {D}ynamics, and
  {N}on-stationarities.
\newblock {\em PLoS Computational Biology}, 10(3), 2014.

\bibitem{Moeller2010}
S.~Moeller, E.~Yacoub, C.~A. Olman, E.~Auerbach, J.~Strupp, N.~Y. Harel, and
  K.~Ugurbil.
\newblock Multiband multislice ge-epi at 7 tesla, with 16-fold acceleration
  using partial parallel imaging with application to high spatial and temporal
  whole-brain {fMRI}.
\newblock {\em Magnetic resonance in medicine}, 63 5:1144--53, 2010.

\bibitem{Pan2010}
S.~Pan and Q.~Yang.
\newblock A survey on transfer learning.
\newblock {\em IEEE Transactions on Knowledge and Data Engineering}, 22:1345 --
  1359, 2010.

\bibitem{Saenger2017}
V.~Saenger, J.~Kahan, T.~Foltynie, K.~Friston, E.~Pereira, A.~Green,
  T.~Van~Hartevelt, J.~Cabral, A.~Stevner, H.~Fernandes, L.~Mancini,
  J.~Thornton, T.~Yousry, P.~Limousin, L.~Zrinzo, M.~Hariz, P.~Marques,
  N.~Sousa, M.~Kringelbach, and G.~Deco.
\newblock Uncovering the underlying mechanisms and whole-brain dynamics of deep
  brain stimulation for parkinson’s disease.
\newblock {\em Scientific Reports}, 2017.

\bibitem{Salimi2014}
G.~Salimi-Khorshidi, G.~Douaud, C.~F. Beckmann, M.~F. Glasser, L.~Griffanti,
  and S.~M. Smith.
\newblock Automatic denoising of functional {MRI} data: Combining independent
  component analysis and hierarchical fusion of classifiers.
\newblock {\em NeuroImage}, 90:449 -- 468, 2014.

\bibitem{Samek2017}
W.~Samek, A.~Binder, G.~Montavon, S.~Lapuschkin, and K.-R. Müller.
\newblock Evaluating the visualization of what a deep neural network has
  learned.
\newblock {\em IEEE Transactions on Neural Networks and Learning Systems},
  28:2660--2673, 2017.

\bibitem{Seabold2010}
S.~Seabold and J.~Perktold.
\newblock Statsmodels: Econometric and statistical modeling with python.
\newblock 2010.

\bibitem{Seth2015}
A.~Seth, A.~Barrett, and L.~Barnett.
\newblock Granger causality analysis in neuroscience and neuroimaging.
\newblock {\em The Journal of neuroscience : the official journal of the
  Society for Neuroscience}, 35:3293--7, 2015.

\bibitem{Seth2012}
A.~Seth, P.~Chorley, and L.~Barnett.
\newblock Granger causality analysis of fmri bold signals is invariant to
  hemodynamic convolution but not downsampling.
\newblock {\em NeuroImage}, 65:540--55, 2012.

\bibitem{Setsompop2012}
K.~Setsompop, B.~Gagoski, J.~R. Polimeni, T.~Witzel, V.~J. Wedeen, and L.~L.
  Wald.
\newblock Blipped-controlled aliasing in parallel imaging for simultaneous
  multislice echo planar imaging with reduced g-factor penalty.
\newblock {\em Magnetic resonance in medicine}, 67 5:1210--24, 2012.

\bibitem{Shuman2013}
D.~I. Shuman, S.~K. Narang, P.~Frossard, A.~Ortega, and P.~Vandergheynst.
\newblock The emerging field of signal processing on graphs: Ex-tending
  high-dimensional data analysis to networks and other irregular domains.
\newblock {\em IEEE Signal Processing Magazine}, 30(3):83--98, 2013.

\bibitem{SmithR2012}
R.~Smith, J.-D. Tournier, F.~Calamante, and A.~Connelly.
\newblock Anatomically-constrained tractography: Improved diffusion mri
  streamlines tractography through effective use of anatomical information.
\newblock {\em NeuroImage}, 62:1924--38, 2012.

\bibitem{SmithR2013}
R.~Smith, J.-D. Tournier, F.~Calamante, and A.~Connelly.
\newblock Sift: Spherical-deconvolution informed filtering of tractograms.
\newblock {\em NeuroImage}, 67:298--312, 2013.

\bibitem{Smith2013}
S.~M. Smith, C.~F. Beckmann, J.~Andersson, E.~J. Auerbach, J.~Bijsterbosch,
  G.~Douaud, E.~Duff, D.~A. Feinberg, L.~Griffanti, M.~P. Harms, M.~Kelly,
  T.~Laumann, K.~L. Miller, S.~Moeller, S.~Petersen, J.~Power,
  G.~Salimi-Khorshidi, A.~Z. Snyder, A.~T. Vu, M.~W. Woolrich, J.~Xu,
  E.~Yacoub, K.~Uğurbil, D.~C.~V. Essen, and M.~F. Glasser.
\newblock Resting-state {fMRI} in the human connectome project.
\newblock {\em NeuroImage}, 80:144 -- 168, 2013.

\bibitem{Sotiropoulos2013b}
S.~Sotiropoulos, S.~Jbabdi, J.~Xu, J.~Andersson, S.~Moeller, E.~Auerbach,
  M.~Glasser, M.~Hernandez~Fernandez, G.~Sapiro, M.~Jenkinson, D.~Feinberg,
  E.~Yacoub, C.~Lenglet, V.~DC, K.~Ugurbil, and T.~Behrens.
\newblock Advances in diffusion mri acquisition and processing in the human
  connectome project.
\newblock {\em NeuroImage}, 80:125, 2013.

\bibitem{Sotiropoulos2013}
S.~Sotiropoulos, S.~Moeller, S.~Jbabdi, J.~Xu, J.~Andersson, E.~Auerbach,
  E.~Yacoub, D.~Feinberg, K.~Setsompop, L.~Wald, T.~Behrens, K.~Ugurbil, and
  C.~Lenglet.
\newblock Effects of image reconstruction on fibre orientation mapping from
  multi-channel diffusion {MRI}: Reducing the noise floor using {SENSE}.
\newblock {\em Magnetic resonance in medicine : official journal of the Society
  of Magnetic Resonance in Medicine / Society of Magnetic Resonance in
  Medicine}, (70), 2013.

\bibitem{Surampudi2018}
S.~G. Surampudi, S.~P.~K. Naik, R.~B. Surampudi, V.~K. Jirsa, A.~Sharma, and
  D.~Roy.
\newblock Multiple kernel learning model for relating structural and functional
  connectivity in the brain.
\newblock {\em Scientific Reports}, 8, 2018.

\bibitem{Sutskever2014}
I.~Sutskever, O.~Vinyals, and Q.~V. Le.
\newblock Sequence to sequence learning with neural networks.
\newblock {\em CoRR}, abs/1409.3215, 2014.

\bibitem{Teng2016}
S.~Teng.
\newblock Scalable algorithms for data and network analysis.
\newblock {\em Foundations and Trends in Theoretical Computer Science},
  12(1-2):1--274, 2016.

\bibitem{Thomas2014}
C.~Thomas, F.~Q. Ye, M.~O. Irfanoglu, P.~D. Modi, K.~S. Saleem, D.~A. Leopold,
  and C.~Pierpaoli.
\newblock Anatomical accuracy of brain connections derived from diffusion {MRI}
  tractography is inherently limited.
\newblock {\em Proceedings of the National Academy of Sciences of the United
  States of America}, 111 46:16574--9, 2014.

\bibitem{Tournier2007}
J.-D. Tournier, F.~Calamante, and A.~Connelly.
\newblock Robust determination of the fibre orientation distribution in
  diffusion mri: Non-negativity constrained super-resolved spherical
  deconvolution.
\newblock {\em NeuroImage}, 35:1459--72, 2007.

\bibitem{Tournier2004}
J.-D. Tournier, F.~Calamante, D.~Gadian, and A.~Connelly.
\newblock Direct estimation of the fiber orientation density function from
  diffusion-weighted mri data using spherical deconvolution.
\newblock {\em NeuroImage}, 23:1176--85, 2004.

\bibitem{Tournier2019}
J.-D. Tournier, R.~Smith, D.~Raffelt, R.~Tabbara, T.~Dhollander, M.~Pietsch,
  D.~Christiaens, B.~Jeurissen, C.-H. Yeh, and A.~Connelly.
\newblock Mrtrix3: A fast, flexible and open software framework for medical
  image processing and visualisation.
\newblock {\em NeuroImage}, 202, 2019.

\bibitem{Tzourio2002}
N.~Tzourio-Mazoyer, B.~Landeau, P.~DF, F.~Crivello, O.~Etard, N.~Delcroix,
  B.~Mazoyer, and J.~Marc.
\newblock Automated anatomical labeling of activations in spm using a
  macroscopic anatomical parcellation of the {MNI MRI} single-subject brain.
\newblock {\em NeuroImage}, 15:273--89, 2002.

\bibitem{Ugurbil2013}
K.~Uğurbil, J.~Xu, E.~Auerbach, S.~Moeller, A.~Vu, J.~Duarte-Carvajalino,
  C.~Lenglet, X.~Wu, S.~Schmitter, P.-F. Van~de Moortele, J.~Strupp, G.~Sapiro,
  F.~De~Martino, D.~Wang, N.~Harel, M.~Garwood, L.~Chen, D.~Feinberg, S.~Smith,
  and E.~Yacoub.
\newblock Pushing spatial and temporal resolution for functional and diffusion
  {MRI} in the human connectome project.
\newblock {\em NeuroImage}, 80, 2013.

\bibitem{VanEssen2013}
D.~Van~Essen, S.~Smith, D.~Barch, T.~Behrens, E.~Yacoub, and K.~Ugurbil.
\newblock The wu-minn human connectome project: an overview.
\newblock {\em NeuroImage}, 80, 2013.

\bibitem{Guldin1998}
O.~J.~G. W.~O. Guldin~1.
\newblock Is there a vestibular cortex?
\newblock {\em Trends in neurosciences}, 21:254--9, 1998.

\bibitem{Wenzel1996}
R.~Wenzel, P.~Bartenstein, M.~Dieterich, A.~Danek, A.~Weindl, S.~Minoshima,
  S.~Ziegler, M.~Schwaiger, and T.~Brandt.
\newblock Deactivation of human visual cortex during involuntary ocular
  oscillations - a pet activation study.
\newblock {\em Brain : a journal of neurology}, 119:101--10, 1996.

\bibitem{Werbos1990}
P.~Werbos.
\newblock Backpropagation through time: what it does and how to do it.
\newblock {\em Proceedings of the IEEE}, 78(10):1550--1560, 1990.

\bibitem{Wirth2018}
A.~Wirth, S.~Frank, M.~Greenlee, and A.~Beer.
\newblock White matter connectivity of the visual-vestibular cortex examined by
  diffusion-weighted imaging.
\newblock {\em Brain Connectivity}, 8:235--244, 2018.

\bibitem{Wu2019}
Z.~{Wu}, S.~{Pan}, F.~{Chen}, G.~{Long}, C.~{Zhang}, and P.~S. {Yu}.
\newblock A comprehensive survey on graph neural networks.
\newblock {\em IEEE Transactions on Neural Networks and Learning Systems},
  pages 1--21, 2020.

\bibitem{Xu2012}
J.~Xu, S.~Moeller, J.~Strupp, E.~Auerbach, L.~Chen, D.~A.~Feinberg, K.~Ugurbil,
  and E.~Yacoub.
\newblock Highly accelerated whole brain imaging using
  aligned-blipped-controlled-aliasing multiband {EPI}.
\newblock {\em Proceedings of the 20th Annual Meeting of ISMRM}, page 2036,
  2012.

\bibitem{Xue2015}
W.~Xue, F.~Bowman, A.~Pileggi, and A.~Mayer.
\newblock A multimodal approach for determining brain networks by jointly
  modeling functional and structural connectivity.
\newblock {\em Frontiers in computational neuroscience}, 9:22, 2015.

\bibitem{Zeiler2013}
M.~Zeiler and R.~Fergus.
\newblock Visualizing and understanding convolutional neural networks.
\newblock {\em ECCV 2014, Part I, LNCS 8689}, 8689, 2013.

\bibitem{Zimmermann2019}
J.~Zimmermann, J.~Griffiths, M.~Schirner, P.~Ritter, and A.~R. McIntosh.
\newblock Subject-specificity of the correlation between large-scale structural
  and functional connectivity.
\newblock {\em Network Neuroscience}, pages 1--35, 2019.

\end{thebibliography}

%%%%%%%%%%%%%%%%%%%%%%%%%%%%%%%%%%
%%%%%%%%%%%%%%%%%%%%%%%%%%%%%%%%%%
%%%%%%%%%%%%%%%%%%%%%%%%%%%%%%%%%%

\clearpage
\section*{Supplementary Material}

%\beginsupplement
\section*{Supplement I} \label{sec:supplement_1}

Here we reproduce the evaluations like described in section \ref{sec:performance} employing a more liberal filtering of the fMRI timecourses. The preprocessing was carried out like described in section \ref{sec:HCP_data} but using a bandpass filter with cutoff frequencies  $0.02 - 0.09 \ Hz$ this time. The training of the two models was performed on 4 resting state fMRI sessions from 40 different subjects. 

\begin{figure}[!htb]
\bc
\makebox[\textwidth][c]
{\includegraphics[width=1.2\textwidth]{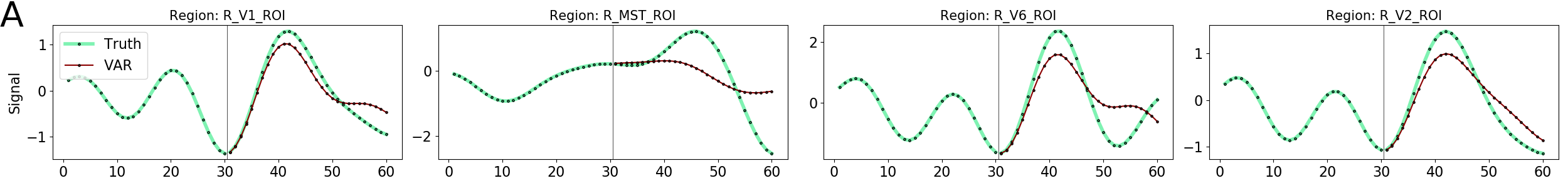}
}

\vspace{4mm}

\makebox[\textwidth][c]
{\includegraphics[width=1.2\textwidth]{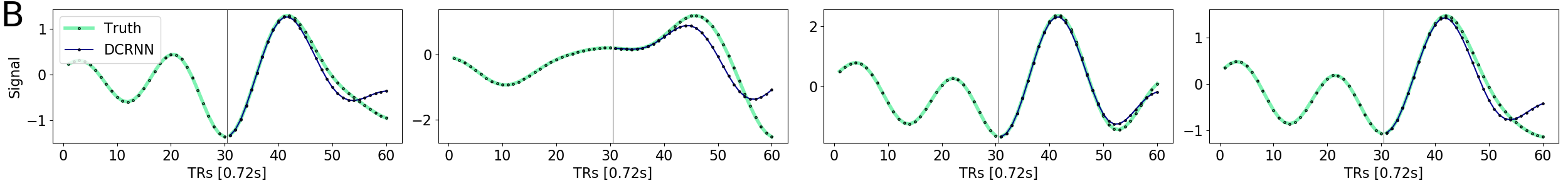}
}

\vspace{1mm}

\makebox[\textwidth][c]
{\includegraphics[width=1\textwidth]{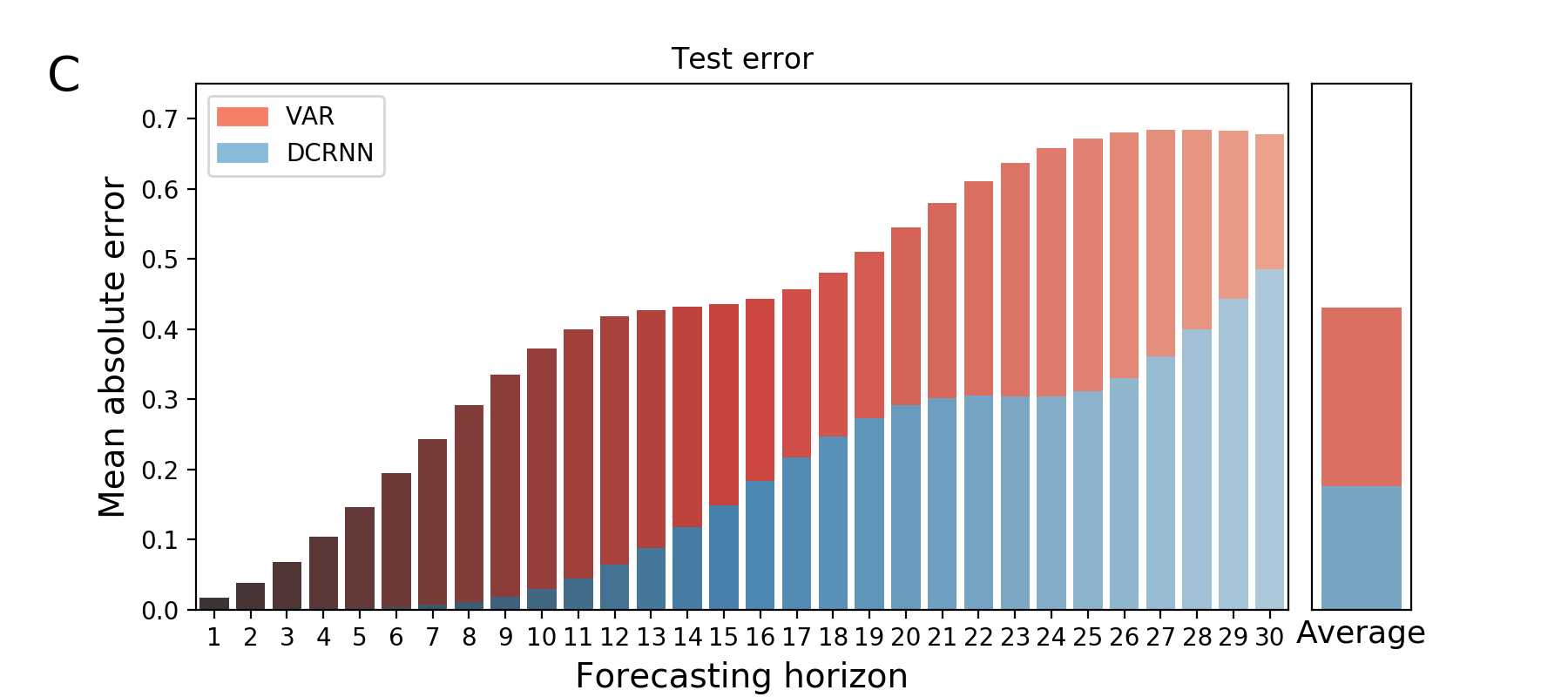}
}
\ec 
\caption{The figure illustrates the prediction accuracy of a VAR model (A) in comparison to the DCRNN (B) in the $0.02 - 0.09 Hz$ frequency range. 
%The true BOLD signal in these 4 ROIs is marked green, while predictions of the VAR are highlighted in red, and for the DCRNN in blue. The first $30 \,\,TRs$ of BOLD signal were used as the model inputs, and the goal was to predict the subsequent $30 \,\,TRs$. 
This illustrative example was chosen to be representative for the whole test set. The prediction error of the VAR model on this sample is with $0.428$ slightly below average, while the error of the DCRNN is with $0.178$ higher than its average. Below the average MAE over all samples in the test set is illustrated, in dependence of the forecasting horizon (C). On the right side in (C) the average of all horizons is shown.}
\label{fig:supp1_comparison}
\end{figure}

\section*{Supplement II} \label{sec:supplement_2}

In this section we replicate the evaluation from section \ref{sec:performance}, using a volumetric parcellation and applying an alternative method for probabilistic tractography of white matter tracks. Volumetric resting-state fMRI images provided by the HCP were subdivided in 90 cortical regions based on the automated anatomical labeling atlas (AAL) \cite{Tzourio2002}. Timecourses within each region were averaged, and like in section \ref{sec:performance}, the $0.04-0.07 Hz$ frequency band was selected for the analysis \cite{Glearean2012}. 
For the reconstruction of anatomical connectivity strengths, the multi-shell ball and stick model \cite{Behrens2007, Jbabdi2012} as implemented in FSL was employed \cite{Jenkinson2012}. Each region of the AAL atlas was defined as seed region, and probabilistic tractography (ProbtrackX) was run based on diffusion parameter estimation with BedpostX \cite{Behrens2007}. For each voxel 5000 samples were generated, and SC was quantified by counting how many streamlines starting in one region reached any other region of the AAL atlas. Those counts were normalized by dividing trough the largest value, and an average SC matrix was computed across 10 different subjects. 

\begin{figure}[!htb]
\bc
\makebox[\textwidth][c]
{\includegraphics[width=1.2\textwidth]{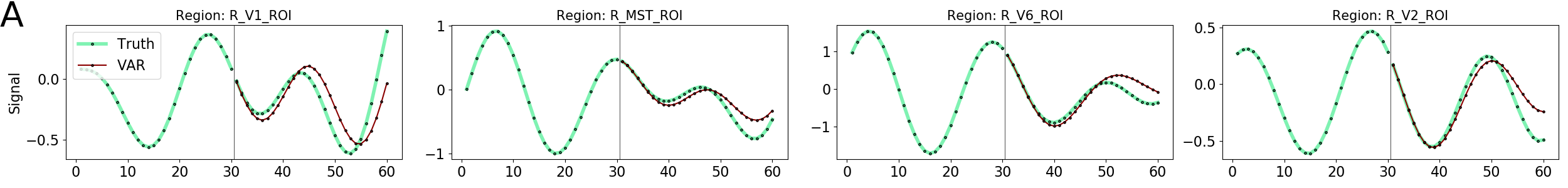}
}

\vspace{4mm}

\makebox[\textwidth][c]
{\includegraphics[width=1.2\textwidth]{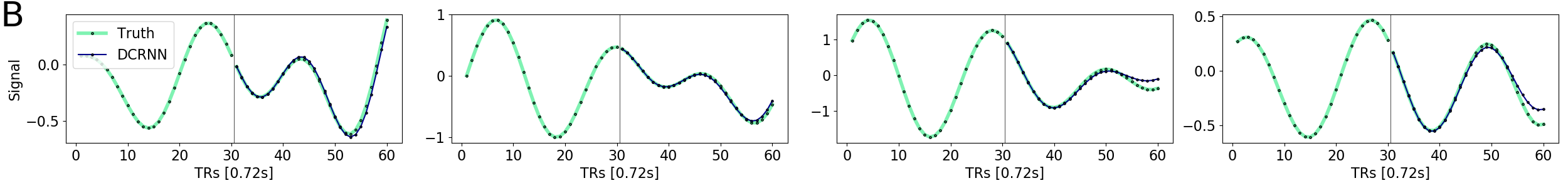}
}

\vspace{1mm}

\makebox[\textwidth][c]
{\includegraphics[width=1\textwidth]{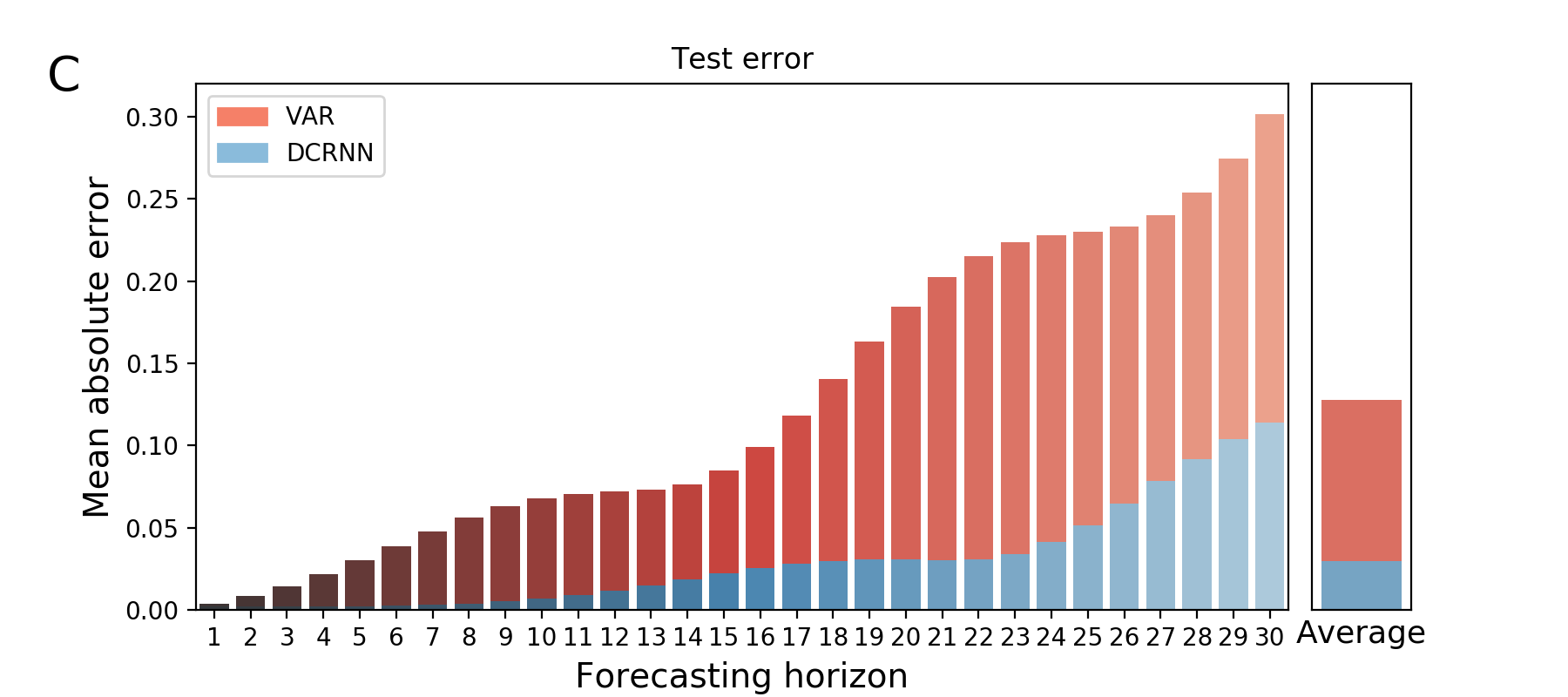}
}
\ec 
\caption{The figure illustrates the prediction accuracy of a VAR model (A) in comparison to the DCRNN (B). 
%The true BOLD signal in these 4 ROIs is marked green, while predictions of the VAR are highlighted in red, and for the DCRNN in blue. The first $30 \,\,TRs$ of BOLD signal were used as the model inputs, and the goal was to predict the subsequent $30 \,\,TRs$. 
This illustrative example was chosen to be reasonable representative for the whole test set, the prediction error of the VAR model on this sample is with $0.119$ slightly below average, while the error of the DCRNN is with $0.033$ higher than its average. Below the average MAE over all samples in the test set is illustrated, in dependence of the forecasting horizon (C). On the right side in (C) the average of all horizons is shown.}
\label{fig:supp2_comparison}
\end{figure}    

\clearpage

\section*{Supplement III} \label{sec:supplement_3}

\begin{figure}[!htb]
\bc
\includegraphics[width=0.8\textwidth]{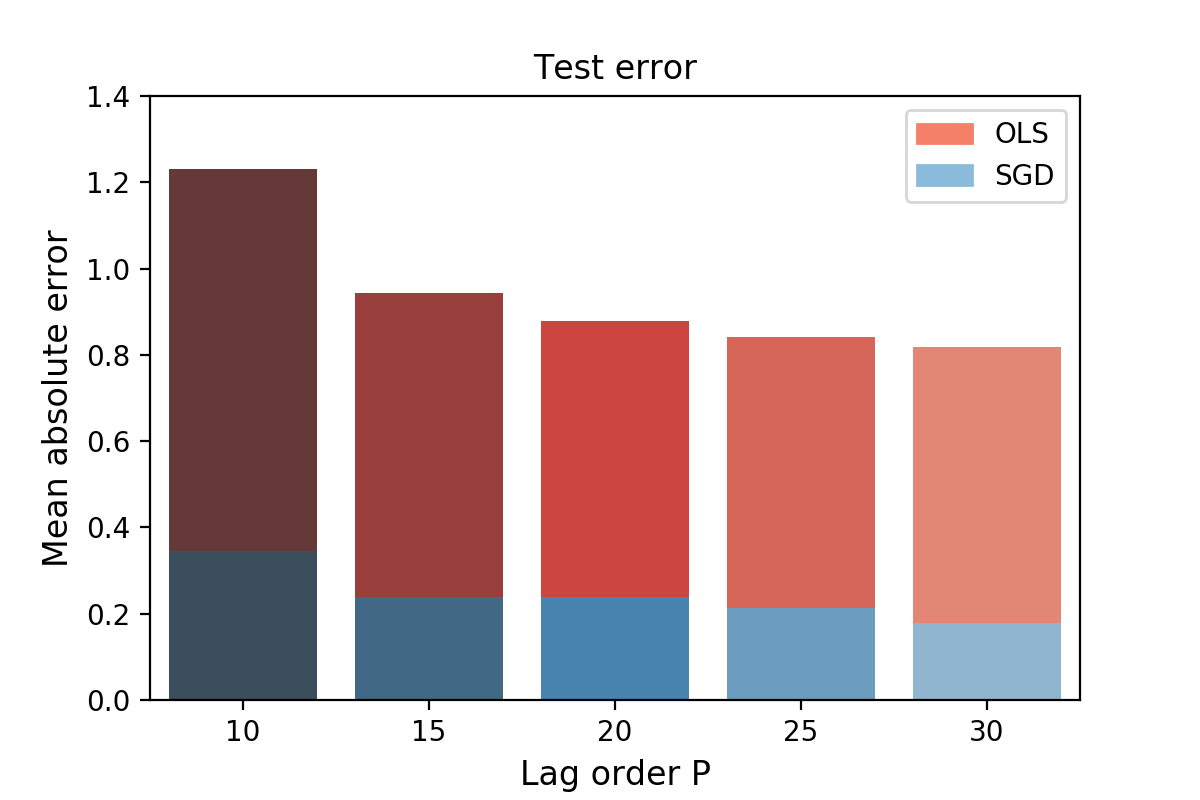}
\ec 
\caption{This figure shows the test MAE for two different optimization strategies to find the VAR coefficients, in dependence of lag orders $P$. The first one is performed with an ordinary least squares (OLS) fit on individual subject sessions and the average test error is depicted in red in this figure. The second one, in analogy to the training of the DCRNN, is based on gradient descent optimization, aggregating input-output pairs of samples across sessions like described in section \ref{sec:data_descr}. The test MAE of the stochastic gradient descent (SGD) approach is illustrated in blue.} 
\label{fig:supp3_comparison}
\end{figure}

%%%%%%%%%%%%%%%%%%%%%%%%%%%%%%%%%%
%%%%%%%%%%%%%%%%%%%%%%%%%%%%%%%%%%
%%%%%%%%%%%%%%%%%%%%%%%%%%%%%%%%%%
\end{document}